\documentclass[namedreferences]{solarphysics}

\usepackage[bookmarks=false, 
     pdfnewwindow=true, 
     colorlinks=true,   
     linkcolor=blue, 
     citecolor=blue,    
     filecolor=blue,    
     urlcolor=blue,   
	final=true]{hyperref}

\usepackage{graphicx}  
\usepackage{array} 
\usepackage{amssymb}        
\usepackage{color}           
\usepackage{breakurl}        
\usepackage{rotating}
\usepackage{longtable} 
\usepackage{pdflscape} 
\usepackage{amsmath}
\usepackage{threeparttable}

\newcommand{\grad}{ {\bf \nabla } }

\newcommand{\aap}{    {\it Astron. Astrophys.}}

\newcommand{\apj}{    {\it Astrophys. J.}}
\newcommand{\apjl}{   {\it Astrophys. J. Lett.}}

\newcommand{\jastp}{  {\it J. Atmos. Solar-Terr. Phys.}} 
\newcommand{\jgr}{    {\it J. Geophys. Res.}}

\newcommand{\solphys}{{\it Solar Phys.}}
 
\newcommand{\ssr}{    {\it Space Sci. Rev.}} 
\chardef\us=`\_

\usepackage[optionalrh, showbiblabels]{spr-sola-addons} 
\begin{document}

\begin{article}
\begin{opening}

\title{Simulating the Coronal Evolution of Bipolar Active Regions to Investigate the Formation of Flux Ropes}

\author[addressref=aff1,corref,email={sly3@st-andrews.ac.uk}]{\inits{S. L.}\fnm{S. L.}~\lnm{Yardley}\orcid{https://orcid.org/0000-0003-2802-4381}}
\author[addressref=aff1]{\inits{D.H.}\fnm{D. H.}~\lnm{Mackay}\orcid{https://orcid.org/0000-0001-6065-8531}}
\author[addressref=aff2]{\inits{L.M.}\fnm{L. M.}~\lnm{Green}\orcid{https://orcid.org/0000-0002-0053-4876}}

\address[id=aff1]{School of Mathematics and Statistics, University of St Andrews, North Haugh, St Andrews, Fife, KY16 9SS, UK}
\address[id=aff2]{Mullard Space Science Laboratory, University College London, Holmbury St. Mary, Dorking, Surrey, RH5 6NT, UK}

\runningauthor{S. L. Yardley {\it et al.}}
\runningtitle{Coronal Evolution of Bipolar ARs}

\begin{abstract}
The coronal magnetic field evolution of 20 bipolar active regions (ARs) is simulated from their emergence to decay using the time-dependent nonlinear force-free field method of \citet{Mackay-2011}. A time sequence of cleaned photospheric line-of-sight magnetograms, that covers the entire evolution of each AR, is used to drive the simulation. A comparison of the simulated coronal magnetic field with the 171 and 193~\AA\ observations obtained by the {\it Solar Dynamics Observatory} (SDO)/{\it Atmospheric Imaging Assembly} (AIA), is made for each AR by manual inspection. The results show that it is possible to reproduce the evolution of the main coronal features such as small- and large-scale coronal loops, filaments and sheared structures for 80\% of the ARs. Varying the boundary and initial conditions, along with the addition of physical effects such as Ohmic diffusion, hyperdiffusion and a horizontal magnetic field injection at the photosphere, improves the match between the observations and simulated coronal evolution by 20\%. The simulations were able to reproduce the build-up to eruption for 50\% of the observed eruptions associated with the ARs. The mean unsigned time difference between the eruptions occurring in the observations compared to the time of eruption onset in the simulations was found to be $\approx$5~hrs. The simulations were particularly successful in capturing the build-up to eruption for all four eruptions that originated from the internal polarity inversion line of the ARs. 
The technique was less successful in reproducing the onset of eruptions that originated from the periphery of ARs and large-scale coronal structures. For these cases global, rather than local, nonlinear force-free field models must be used. While the technique has shown some success, eruptions that occur in quick succession are difficult to reproduce by this method and future iterations of the model need to address this.


\end{abstract}
\end{opening}

\section{Introduction}
     \label{sec:intro} 

The solar corona is highly complex in nature. The source of its complexity is largely due to the presence of magnetic fields that are generated in the tachocline \citep{Spiegel-1992}: a region close to the base of the convection zone \citep{Charbonneau-2010, Charbonneau-2014}. When magnetic flux tubes at the base of the convection zone become unstable to buoyancy \citep{Parker-1955, Zwaan-1985} they rise and the magnetic field breaks through the solar surface manifesting itself as an active region (AR) in the photosphere. The magnetic flux emerges in a non-potential state \citep{Leka-1996} and is further modified by the action of photospheric flows. This results in free magnetic energy being available to drive solar eruptive phenomena. 

ARs are the source of a wide range of atmospheric solar activity and the type and level of activity is dependent on the evolutionary stage of the AR (for a review on AR evolution see \citealt{vanDriel-2015}). As a result, it is important to understand the structure and evolution of the magnetic field of an AR over its entire lifetime, from emergence to decay. 

It is currently difficult to measure the magnetic field in the corona and extreme ultraviolet (EUV) observations of AR coronal loops can only provide indirect and limited information of the coronal structure of ARs. An alternative approach, for the analysis of the coronal structure of ARs, is to construct a model of the coronal magnetic field by using the photospheric magnetic field as the lower boundary condition. This approach relies on the approximation that the corona, a low plasma-$\beta$ environment that mostly remains in equilibrium, is ``force-free". This means that the coronal magnetic field must satisfy the criterion of $\boldsymbol{j} \times \boldsymbol{B} = 0$ where $\boldsymbol{j} = \alpha \boldsymbol{B}$. In the case of nonlinear force-free (NLFF) fields the torsion parameter $\alpha$ is a scalar function that can vary as a function of position, but must remain constant along magnetic field lines. 


There are numerous NLFF field techniques that can be used to generate models of the coronal magnetic field. These NLFF field models can be divided into two categories: models that are static or time-dependent. Static models either use a vector magnetogram as the lower boundary condition and extrapolate the NLFF fields into the corona (e.g. \citealt{Schrijver-2006, DeRosa-2009, Canou-2010, Wiegelmann-2012, Jiang-2014}), or they take an initial coronal field, which is either a potential or linear force-free (LFF), and evolve this field into a NLFF state. The latter approach can make use of the magnetofrictional relaxation technique \citep{Yang-1986} to generate a static model of the magnetic field of an AR. Examples of static modelling using magnetofrictional relaxation include the magnetofrictional extrapolation method of \citet{Valori-2005} and the flux rope insertion method \citep{vB-2004, Bobra-2008, Savcheva-2012, Yardley-2019}. The extrapolation methods mentioned above produce a coronal field model at a single snapshot in time. A series of independent, static extrapolations may be produced but there is no direct evolution from one extrapolation to the next.

The magnetofrictional relaxation technique can also be used as a simulation method to construct a continuous time-dependent series of NLFF fields. In this case, the normal component of the magnetic field is specified along with an initial field and a time series of horizontal boundary motions. The resulting coronal structures are due to the applied boundary motions injecting non-potentiality into the corona over timescales of hours or days. The coronal field, which is in non-equilibrium is then relaxed back to a NLFF field equilibrium using magnetofrictional relaxation. This has been applied to global simulations \citep{Mackay-2006a, Mackay-2006b} where a flux transport model is applied at the photospheric boundary or to simulate AR evolution using a time series of line-of-sight (LoS) magetograms \citep{Mackay-2011, Gibb-2014} or more recently vector magnetograms (e.g. \citealt{Pomoell-2019}). 


In the recent study by \citet{Yardley-2018b} a continuous time-dependent series of NLFF field models of AR 11437 were created using the time-dependent NLFF field method of \citet{Mackay-2011}. Photospheric LoS magnetograms from the SDO/{\it Helioseismic Magnetic Imager} (HMI) instrument were used as lower boundary conditions to drive the simulation and continuously evolve the coronal field through a series of NLFF equilibria. When the results from the simulation were compared to SDO/AIA observations it was found that the simulation was able to capture the majority of the characteristics of the coronal field evolution. Flux ropes that formed in the simulation showed signatures of eruption onset for two out of three of the observed eruptions, approximately 1 and 10~hrs before the eruptions occurred in the observations. A parameter study was also conducted to test whether varying the initial condition and boundary conditions along with the inclusion of Ohmic diffusion, hyperdiffusion, and an additional horizontal magnetic field injection at the photosphere affect the coronal evolution and timings of the eruption onset. The results showed that the coronal evolution and timings of eruption onset were not significantly changed by these variations and inclusions, indicating that the main element in replicating the coronal field evolution is the Poynting flux from the boundary evolution of the LoS magnetograms. AR 11437 is also included in this current study.




In this paper, we extend the set of simulations carried out in \citet{Yardley-2018b} of a single AR by simulating the coronal magnetic field evolution of 20 bipolar ARs. The observational analysis of the same set of bipolar ARs was conducted by \citet{Yardley-2018a} in order to probe the role of flux cancellation as an eruption trigger mechanism. The study of \citet{Yardley-2018a} analysed both photospheric and coronal observations taken by SDO over the entire lifetime of the ARs. Through simulating a much larger sample of ARs we can obtain more general results than those found in \citet{Yardley-2018b}, which only considered a single region (AR 11437). We aim to determine whether the simulation of a series of NLFF fields using the magnetofrictional technique can capture the coronal evolution and also the build-up phase that brings the coronal field to the point of eruption. The analysis carried out here is similar to that of \citet{Yardley-2018b} in which the NLFF field method was tested. However, due to the large-scale analysis of 20 ARs the results are presented in less detail than those given in \citet{Yardley-2018b}.

The outline of the paper is as follows. Section~\ref{sec:obs} outlines the observations including the criteria for AR selection, coronal evolution and eruptions produced by each AR. Section~\ref{sec:sim} describes the technique used to simulate the coronal field including the lower boundary conditions used. Results from the simulations can be found in Section~\ref{sec:res}, which includes simulations using the simplest initial and boundary conditions and also with the inclusion of additional effects. Section~\ref{sec:dis} discusses the results and Section~\ref{sec:sum} provides a conclusion to the study.

\section{Observations} \label{sec:obs}

\subsection{AR Selection}

The 20 ARs presented in \citet{Yardley-2018a} are the same regions used in this study. We now briefly summarise the data selection method used by \citet{Yardley-2018a} to identify and select these ARs and refer the reader to that paper for more details on each region. ARs were selected using the following criteria:

\begin{enumerate}
    \item The ARs must be bipolar and have low complexity. The regions must have two dominant photospheric magnetic polarities with no major mixing of the opposite polarities.
    
    \item The ARs must be isolated with minimal interaction occurring between the AR and other ARs or the background quiet Sun magnetic field.
    
    \item The ARs must be observable from their first emergence and form east of central meridian. This allows the full evolution from emergence to decay to be simulated during disk transit.
    
    \item The ARs first emergence must be no more than 60$^{\circ}$ from central meridian as instrumental effects become increasingly significant at large centre-to-limb angles.
\end{enumerate}

These selection criteria led to a sample of 20 ARs being chosen during the HMI era, spanning a time period from March 2012 to November 2015. All ARs, apart from AR 11867, were monitored during their flux emergence and decay phases, which included dispersal and flux cancellation. AR 11867 remained in its emergence phase during the time period studied and did not exhibit flux cancellation at its internal PIL.

\begin{figure}
\centerline{\includegraphics[width=1\textwidth,clip=]{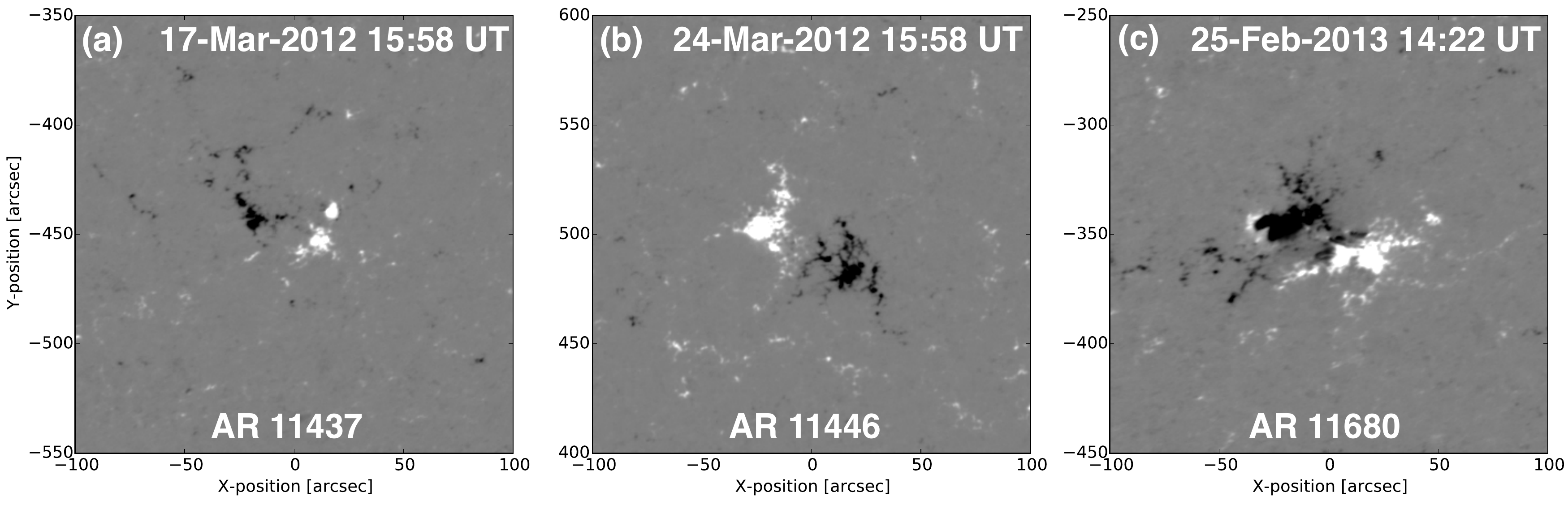}}
\caption{SDO/HMI LoS magnetograms that show three AR examples (ARs 11437, 11446 \& 11680). The images show each AR at the time of the peak unsigned magnetic flux measurement, where unsigned refers to half the total absolute positive and negative flux. The saturation levels of the images are $\pm$~100~G with white (black) representing positive (negative) photospheric magnetic field. As an example, the entire photospheric field evolution of AR 11437 can be seen online in Supplementary Movie 1.}
\label{fig1}
\end{figure}

Representative AR examples are given in Figure~\ref{fig1} with Supplementary Movie 1 showing the full evolution of AR 11446. Table~\ref{tab:table1} provides summary information of AR locations, photospheric flux evolution, and observed eruption times taken from \citet{Yardley-2018a}. Photospheric flux values were obtained using the 720~s data series \citep{Couvidat-2016} generated by the {\it Helioseismic Magnetic Imager} (HMI) \citep{Schou-2012} on board the {\it Solar Dynamics Observatory} (SDO); \citealt{Pesnell-2012}).

\subsection{Coronal Evolution and Eruptive Activity}

The observed coronal evolution of each AR was analysed in \citet{Yardley-2018a} in order to identify the time and location of any eruptions. These ejections are referred to as eruptions as opposed to CMEs because the coronal signatures in the EUV data are relatively subtle and most do not show any clear evidence of a CME in the white-light coronagraph data. This implies that they are either confined/failed eruptions or are ejective but have a low plasma density. The coronal evolution was monitored using both 171 and 193~\AA\ images taken by the {\it Atmospheric Imaging Assembly} (AIA; \citealt{Lemen-2012}) on board SDO. AIA provides full-disk observations with a high spatial and temporal resolution of 1.5" and 12~s, respectively. At least two or more of the following coronal signatures were used to identify the occurrence of an eruption:

\begin{enumerate}
    \item the eruption of a filament or an EUV loop system,
    \item the rapid disappearance of coronal loops and post-eruption arcade formation (flare arcade),
    \item flares and flare ribbons,
    \item and/or coronal dimmings.
    
\end{enumerate}

As detailed in \citet{Yardley-2018a} the eruptions were then categorized into the following types to investigate which eruptive structures might have formed as a consequence of flux cancellation:

\begin{enumerate}
    \item Internal PIL events are the eruption of a low altitude structure originating along the internal PIL of the AR.
    
    \item External PIL events are the eruption of a low altitude structure originating along an external PIL that is formed between the periphery of the AR and the magnetic field of the quiet Sun.
    
    \item High altitude events are the eruption of a high altitude structure which cannot be associated with an internal/ external PIL (which are at low altitude).
\end{enumerate}

In total, 24 eruptions were observed, with 13 of the 20 ARs producing at least one ejection. Eight of these ARs produced low corona events originating from either the internal or external PIL and the other five produced high altitude events. Two of the eruptions were observed as a CME in the LASCO/C2 coronagraph data. There were also four B/C GOES class flares associated with four ARs that did not occur at the time of the eruptions.

For examples of the different event categories see Figure~1 in \citet{Yardley-2018a}. The timings of these events, which are also taken from \citet{Yardley-2018a}, are given in Table~\ref{tab:table1}.

\section{The NLFF Field Simulation} \label{sec:sim}

\subsection{Coronal Magnetic Field Evolution}
The NLFF field method of \citet{Mackay-2011} is applied to SDO/HMI LoS magnetograms to simulate the evolution of the coronal magnetic field of each AR. A key element of this method is that the magnetic field evolves through a continuous time series both at the photosphere and in the coronal volume where flux is preserved. Therefore, the coronal magnetic field evolution can be analysed. When using our method we do not apply any additional observational constraints such as the use of EUV coronal images, rather the solution obtained at any one time is purely based on the initial field, the applied boundary motions and any additional coronal physics (see Section~\ref{sec:par}).

This technique has been previously tested on AR 11437 \citep{Yardley-2018b}, one of the ARs also included in this study. Therefore, the quantitative analysis that has previously been carried out for AR 11437 will not be described in this paper. Here, we present the overarching results from the qualitative analysis of 20 bipolar ARs, where each AR has been studied using the methodology described in \citet{Yardley-2018b}.

A time series of NLFF fields is generated using HMI LoS magnetograms for each lower boundary condition (see Section~\ref{sec:bc}). The HMI LoS magnetograms are cleaned and re-scaled before the simulations are carried out. The clean-up procedure includes time-averaging, low magnetic flux value removal, removal of small-scale magnetic elements, and if required, flux balancing. This procedure ensures that the large-scale AR evolution is kept but small-scale quiet Sun elements and random noise are removed (see Appendix~\ref{sec:AppA} for more details).

In the simulation, the evolution of the 3D magnetic field $\boldsymbol{B}$ is described by
\begin{equation}
\frac{\partial \boldsymbol{A}}{\partial t} = \boldsymbol{v} \times \boldsymbol{B}, \label{eq:ind}
\end{equation}
where $\boldsymbol{A}$ represents the magnetic vector potential, $\boldsymbol{B} = \nabla \times \boldsymbol{A}$ is the magnetic field, and $\boldsymbol{v}$ is the magnetofrictional velocity. The magnetofrictional relaxation technique of \citet{Yang-1986} is employed to ensure that the coronal field is evolved through a series of force-free equilibria. Therefore, the magnetofrictional velocity inside the computational box takes the form
\begin{equation}
\boldsymbol{v} = \frac{1}{\nu'}  \boldsymbol{j} \times \boldsymbol{B},
\end{equation}
where $\boldsymbol{\nu'}$ is the friction coefficient and $\boldsymbol{j} = \nabla \times \boldsymbol{B}$. The coefficient of friction ensures that as the magnetic field is perturbed by motions at the boundary, the field remains close to a force-free equilibrium in the corona. A cartesian staggered grid is used to carry out the computations to obtain second-order accuracy for $\boldsymbol{A}$, $\boldsymbol{B}$, and $\boldsymbol{j}$. The computational domain represents the solar corona where the photosphere is represented by the bottom of the box. The size of the computational domain ranges from $0 < x, y, z < 6$ in non-dimensionalized units, where the size of the computational box in physical units is on the order of 10$^{5}$~km. The exact domain size depends upon the dimensions of the original magnetograms and how the magnetograms are re-scaled within the computational box (see Section \ref{sec:bc}). 
The sides of the computational domain have closed boundary conditions whereas, the top of the box can have either open or closed boundaries. When the top of the computational domain is open then the magnetograms do not need to be flux balanced. However, when the top of the box is closed then the magnetograms require flux balancing to ensure that $\nabla \cdot \boldsymbol{B} = 0$ in the computational volume. In this particular study, both open and closed boundary conditions are used for the top of the box. The generation of the photospheric boundary and initial conditions are described below.

\subsection{Photospheric Boundary Conditions} \label{sec:bc}

To be able to simulate the full evolution of the bipolar ARs we use the full disk HMI 720s LoS magnetograms (hmi.M\_720s series). For each AR, we use a time sequence of LoS magnetograms with a chosen cadence of 96 minutes. We create cut-outs of the magnetograms centred on each AR and apply clean-up processes to the time series of partial disk magnetograms (see Appendix~\ref{sec:AppA}). We use LoS magnetograms in this study as we want to simulate the full evolution of ARs from emergence to decay. We would also like to quantify how well this computationally efficient modelling technique that uses the LoS magnetograms performs in simulating the coronal evolution of a large number of ARs. 

Regarding the medium cadence used, prior to the present study, we have conducted a number of investigations varying the cadence of the HMI magnetograms from 12 minutes to 3 hours \citep{Gibb-2015} and have found very similar results. Therefore, we have chosen to use a medium cadence of 96 minutes as it is sufficient to capture the large-scale evolution of the ARs.
Also, any future L5 space weather mission is likely to have a cadence more comparable to that of the Michelson Doppler Imager (MDI) rather than the cadence presently provided by HMI.  


Initially, each simulation is run using a relatively simple set-up. That is, a potential magnetic field is used as the initial condition along with either a closed or open boundary at the top of the computational volume. The simulation results are then compared with the observations to determine whether there is a good agreement between the two. This is assessed by comparing the evolution of the simulated coronal field to the coronal evolution in SDO/AIA 171 and 193~\AA\ observations by visual inspection and using qualitative scoring criteria given in Section~\ref{sec:simp}. If the simulation results do not provide a good fit to the observations then the simulation is re-run varying a number of terms one-by-one. First a LFF field initial condition is used then a variety of additional physical effects are included in succession until a better fit is achieved (see Section~\ref{sec:par} and also the method of \citealt{Yardley-2018b}). For the present simulations we only use a potential or a LFF field as the initial condition for the simulations. The ARs modelled in this study are young ARs, with the majority emerging at a centre-to-limb angle around 60$^{\circ}$ longitude. Due to the large distance from central meridian the vector magnetograms (where they may exist) contain significant errors where these errors could introduce spurious results in the simulation. Therefore, using an initial NLFF field condition is currently beyond the scope of this paper but this will be considered in a future study.

The simulations use the cleaned LoS magnetograms (see Appendix~\ref{sec:AppA}), which have a been scaled to a lower resolution of 256$^{2}$, as the lower boundary conditions. The original size of the magnetograms depends upon the size of the AR but the LoS magnetograms are always larger than 256$^{2}$. To take into account boundary effects, the magnetograms are also re-scaled to fill 60--70\% of the area of the bottom of the computational box. The simulation generates a continuous series of lower boundary conditions using the corrected LoS magnetograms that are designed to replicate, pixel by pixel the LoS magnetograms, every 96~minutes.


The series of cleaned magnetograms give the prescribed distribution of $B_{z}$ on the base. Hence the horizontal components of the vector potential ($A_{xb}, A_{yb}$) are determined on the base for each discrete time interval of 96 minutes by solving for the scalar potential $\phi$, where $\boldsymbol{A} = \nabla \times (\phi \boldsymbol{\hat{z}}) $. To specify the evolution of $B_{z}$ on the base in terms of $A_{xb}$ and $A_{yb}$ between the prescribed distributions the rate of change of the horizontal components of the magnetic vector potential and therefore an electric field is determined. To evolve $A_{xb} (t)$ and $A_{yb} (t)$ to $A_{xb} (t+1)$ and $A_{yb} (t+1)$ we assume that the process is linearly applied between each discrete time interval $t$ and $t+1$,
where $t$ represents the discrete 96 minute time index. Therefore, the horizontal components ($A_{xb}, A_{yb}$) are linearly interpolated between each 96 minute time interval to produce a time sequence that is continuous between the observed distributions. Thus, every 96 minutes the simulated photospheric field identically matches that found in the cleaned observations. By using this technique,  we are effectively evolving the magnetic field from one fixed magnetogram to the next. Also, undesirable effects such as the pile-up of magnetic flux at sites of flux cancellation and numerical overshoot do not occur. As the surface field evolves in this manner it injects electric currents and free energy into the coronal field, which responds through Equation \ref{eq:ind}. By using this numerical method it means that there are two timescales involved in the lower boundary condition evolution. The first timescale is due to the 96~minute time cadence of the observations and the second is the linear evolution timescale. The second timescale is introduced to advect the photospheric magnetic polarities between the observed states, inject Poynting flux into the corona and to relax the coronal field. The method applied to interpolate the boundary magnetic field is very similar to \citet{Gibb-2014} and \citet{Yardley-2018b} however, to satisfy the Courant-Friedrichs-Levy (CFL) condition the timestep is determined from the minimum cell crossing time for the magnetofrictional velocity or the diffusion terms and its maximum is equal to a fifth of this value.

Within the simulations the initial condition satisfies the Coulomb gauge. In addition to this, during the evolution of the field between the fixed points given by the magnetograms, we also maintain the Coulomb gauge. This is carried out numerically by including a $\grad \cdot \mathbf{A}$ term, which does not effect the value of the magnetic field in the simulations. The complete description of this process can be found in \citet{Mackay-2009}, \citet{Mackay-2011} and references therein.

\section{Results} \label{sec:res}

\subsection{Magnetic Field Evolution} \label{sec:simp}

\begin{figure}
\centerline{\includegraphics[width=1\textwidth,clip=]{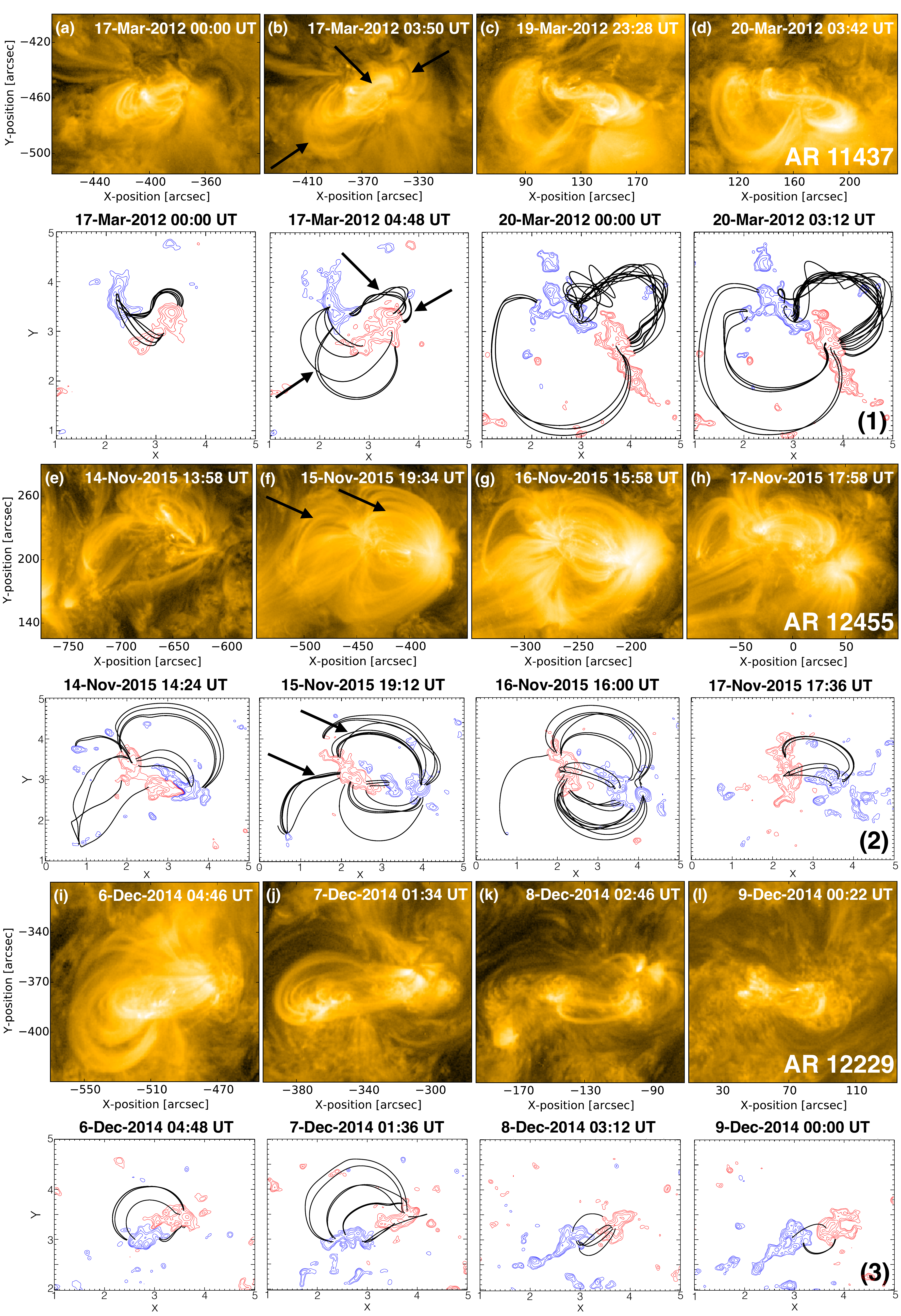}}
\caption{Example ARs shown for each scoring criteria (1, 2, and 3). The odd rows (1, 3, and 5) show the evolution of the ARs through {\it SDO}/AIA 171~\AA\ images, where the NOAA AR number is labelled on the image in the final column (panels d, h, and l). The even rows (2, 4, and 6) show the corresponding simulated coronal field at the time of the coronal observation shown directly above, where the score is labelled on the simulated coronal field in the final column. The red (blue) contours shown in the images of the simulated coronal field evolution represent the positive (negative) photospheric magnetic field polarities. The black arrows indicate the observed coronal features such as sheared structures, filaments and small- and large-scale loops, which are reproduced by the simulations. }
\label{fig2}
\end{figure}

The simulated coronal field evolution of the 20 bipolar ARs will now be discussed for the simplest case where a potential field is used as the initial condition and the top boundary of the computational box is closed. To determine whether the simulated coronal evolution is able to capture that of the real Sun in each AR, the simulated field is compared to the SDO/AIA 171 and 193~\AA\ plasma emission structures by manual inspection. The main coronal features that are used to make the comparison between the observed coronal structure and simulated coronal magnetic field of each AR include small- and large-scale coronal loops, filaments and sheared structures. The 171 and 193~\AA\ wavebands are used for the comparison as the evolution of coronal loops, filaments and sheared structures are well captured in these wavebands compared to the other AIA wavebands. These wavebands are also the primary wavebands that were analysed in the observational study of \citet{Yardley-2018a}. The simulated magnetic field and observed coronal plasma emission structures are then compared at various times (roughly once per day, see Figure~\ref{fig2}) throughout the evolution of each AR. 


The simulation results are also analysed to determine whether or not there is a good agreement between the timings and location of the ejections seen in the observations and the corresponding signatures of eruption onset in the simulations. The following criteria are used to assign a score to quantitatively describe the level of agreement between the simulations and observations:

\begin{enumerate}
    \item Score 1: If the simulation is able to reproduce the main coronal features (small- and large-scale loops, filaments and sheared structures) for the majority of the AR evolution then there is deemed to be a good match between the observations and simulations. If an eruption is observed to originate from the AR, then the simulation must be able to successfully model the build-up to the eruption within a $\pm$12~hr time window pre- or post-observed eruption time. If there are multiple observed eruptions then the simulation must be able to successfully follow the build-up to eruption for the majority of the eruptions associated with the AR.

    \item Score 2: Some of the coronal features (small- and large-scale loops, filaments and sheared structures) that are seen in the observations are reproduced by the simulation for most of the AR evolution. Therefore, the match between the coronal features present in the observations and the simulations is deemed to be acceptable. If one or multiple eruptions are observed to originate from the AR, the build-up phase may or may not be followed by the simulation for any eruption.
    
    \item Score 3: A minority or none of the coronal features (small- and large-scale loops, filaments and sheared structures) seen in the observations are reproduced for most of the AR evolution. Therefore, the evolution of the simulated coronal field is deemed not to match the observed coronal evolution. The simulation fails to model the build-up to eruption for any observed eruptions associated with the AR.
    
\end{enumerate}


An example AR for each of the scoring criteria is shown in Figure~\ref{fig2}, which compares the observed coronal evolution (odd rows) to the simulated coronal evolution (even rows). The first example shows AR 11437 (Score 1), where the sheared J-shaped structure, small- and large-scale coronal loops that are present in the observations are captured by the simulation for the majority of the AR evolution (see black arrows in Figure~\ref{fig2}). The simulation is also able to replicate the build-up to the point of eruption for 2 out of 3 of the observed eruptions. The signatures of eruption onset in the simulations are discussed in the next paragraph. The second example shows AR 12455 (Score 2), where the simulation is able to reproduce the structure of the small- and large-scale coronal loops although, the match to the observations is better in the northern part of the AR compared to the south (black arrows in Figure~\ref{fig2}). There are no eruptions observed to be associated with this AR. Finally, for AR 12229 (Score 3) the simulation is unable to produce the structure of the small- and large-scale loops seen in the observations of the AR. The eruption onset signatures, which indicate that a loss of equilibrium in the simulation has occurred, are not present for any of the four eruptions observed to originate from this AR.


The simulations carried out, focus on modelling the build-up of non-potential magnetic fields and flux ropes within ARs. We do not try to reproduce and follow the dynamics of the observed eruptions as full magnetohydrodynamic (MHD) simulations are required to do this (e.g. see \citealt{Rodkin-2017}). Therefore, to determine whether the simulations successfully follow the build-up to eruption, the simulated coronal field evolution was examined for signatures of eruption onset. The signatures present in the simulations that indicate the build-up to an eruption include:

\begin{enumerate}
    \item a flux rope rising, which subsequently reaches the top or side boundaries of the computational box indicating that a loss of equilibrium has occurred.
    
    \item Reconnection occurring underneath the flux rope which leads to small, more potential loops forming beneath the flux rope similar to the post-eruption (flare) arcades that are visible in the observations.
    
\end{enumerate}

\begin{figure}
\centerline{\includegraphics[width=1\textwidth,clip=]{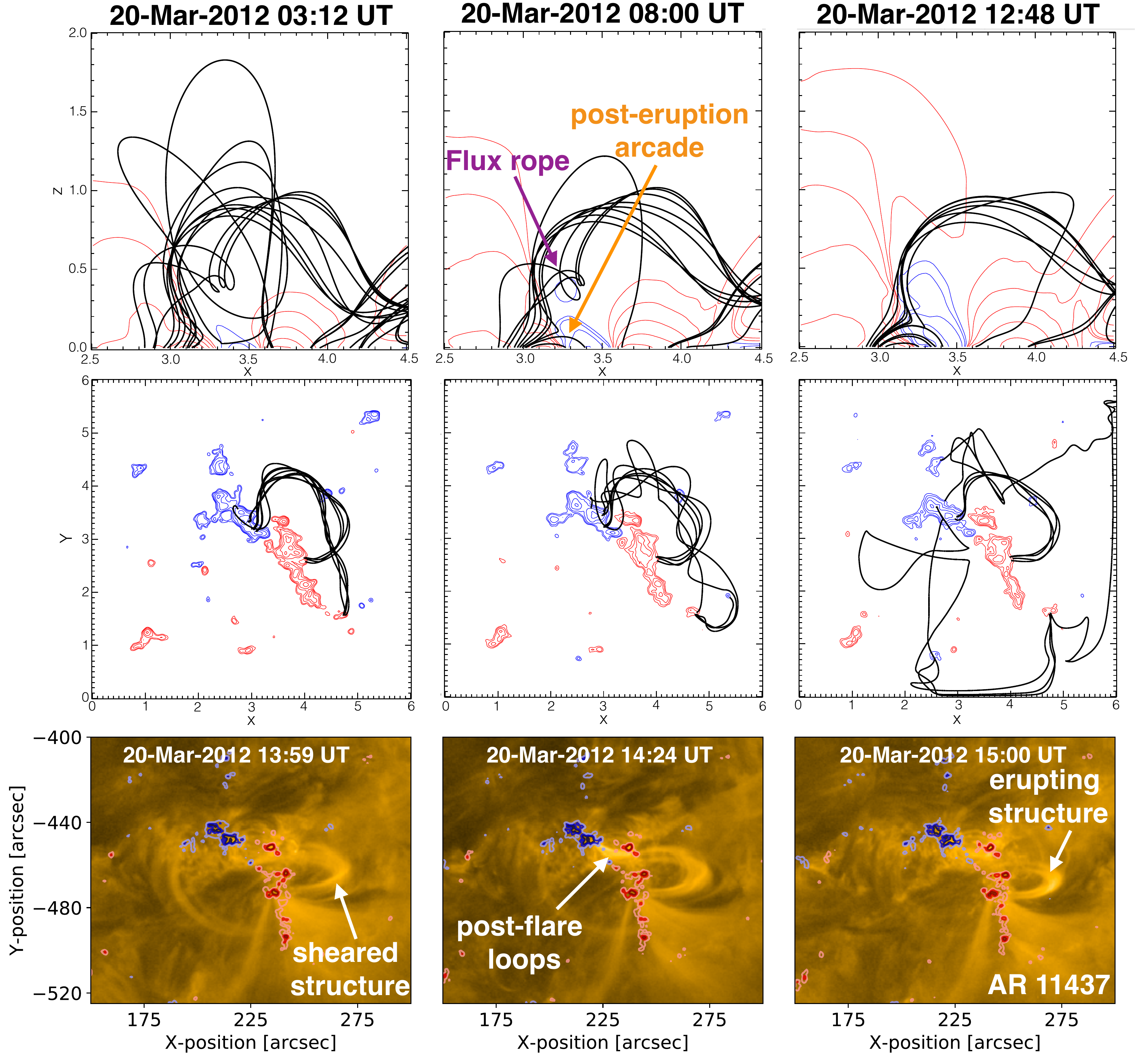}}
\caption{
Sample field lines of AR 11437 that show the presence of a flux rope in the simulation in the build-up to eruption in comparison to the eruption signatures in the observations. The series of plots in the top panel show a cross-section in the x-z plane where the axis of a flux rope and a post-eruption arcade are visible. The middle panel shows the eruption onset of the flux rope in the x-y plane. In this case, the side and top boundaries of the computational box are closed and so the flux rope is unable to escape the coronal volume however, the flux rope does reach the side boundaries of the domain. The bottom panel shows the eruption signatures present in the 171~\AA\ observations.  The red and blue contours represent the positive and negative photospheric magnetic polarities, respectively.}
\label{fig3}
\end{figure}

These signatures of eruption onset in the simulations must occur at the same location and timings as those identified in the observations. The simulation results are analyzed in a time window of $\approx$12~hrs pre- and post-observed eruption for the above signatures of eruption onset.
The signatures of eruption onset in the simulation of AR 11437 are shown in Figure~\ref{fig3}. In this case, a flux rope, which has formed along the internal PIL, rises in the domain and reconnection occurs underneath the flux rope. This leads to small, more potential loops forming below the flux rope axis. Eventually the flux rope reaches the side boundary of the domain. A similar scenario is seen in the observations where a sheared structure and post-eruption loops that form underneath this structure are observed at the same location as in the simulations.

For the simplest case, where the coronal evolution of each of the 20 bipolar ARs is simulated using a potential field initial condition and a closed top boundary, the results (see Table~\ref{tab:table2}) are as follows. The NLFF field method is able to capture the majority of the coronal structure for ten ARs, a reasonable amount of the structure for six ARs, and little or no structure for four ARs (see Table~\ref{tab:table3}). Therefore, the method is able to capture a reasonable amount of the structure for 80\% of the AR sample, and failed to capture the structure for 20\% of the ARs.

\begin{figure}
\centerline{\includegraphics[width=1\textwidth,clip=]{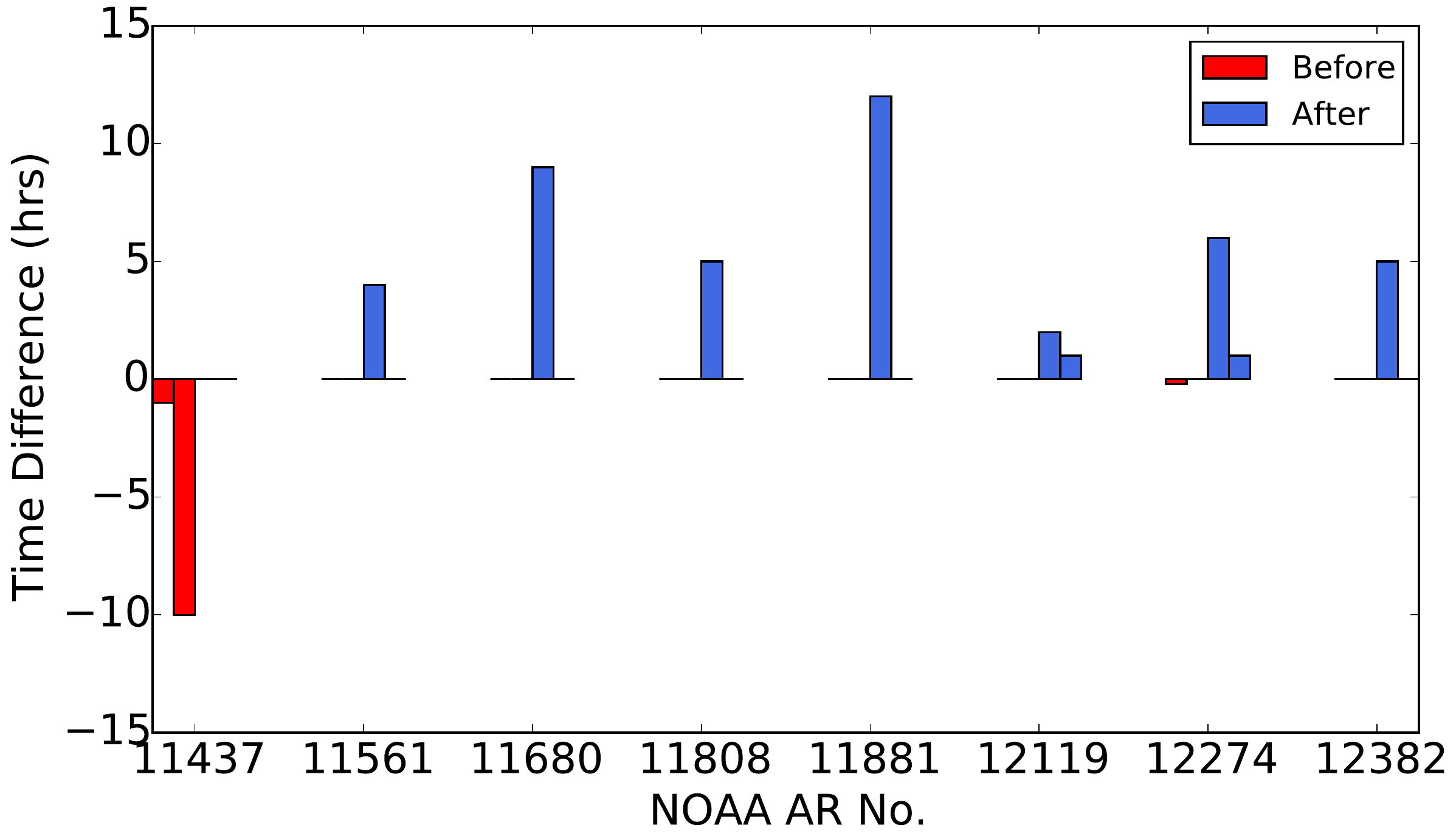}}
\caption{The time difference in hours between the eruption onset of flux ropes in the simulation of each AR compared to the eruption seen in the observations. Signatures of eruption onset occur between two time steps in the simulation and so the values represent the time difference taken between the observed eruptions and the central time between the two time steps in the simulation. }
\label{fig4}
\end{figure}

In total, the simulations are able to successfully follow the build-up to eruption in a $\approx$12~hr time window prior to or post-eruption for 12 out of the 24 observed eruptions. The time difference between eruption onset in the simulations compared to the time determined from observations is given in Figure~\ref{fig4} for each AR. The time of eruption onset in the simulation is determined by using the time halfway between the time step where the signatures of eruption onset in the simulation have been identified, and the previous time step where there are no signatures of eruption onset. By time step we are referring to the primary timescale of the simulation that is set by the cadence of the magnetograms, which in this case is 96~minutes. The time of the eruption onset identified from the simulation is then compared to the eruption time taken from the observations to give the time difference. The mean time difference between the initiation of the eruption in the simulations compared to the observations is $\approx$5~hrs with a standard deviation of $\approx$4~hrs. It is possible to successfully follow the build-up to eruption in the simulations for all four eruptions (100\%), that were observed to originate from low in the corona along the internal PIL by \citet{Yardley-2018a}.

This indicates that by applying the method of \citet{Mackay-2011} to construct a time series of NLFF fields, using the simplest initial and boundary conditions, it is possible to capture the key features of the observable coronal structures in the sample of ARs. To improve on these results the effect on the simulated coronal magnetic field of additional physical effects as well as varying the initial and boundary conditions are examined in the following section.

\subsection{Consequences of Additional Physical Effects} \label{sec:par}

Although it is possible to simulate the coronal field evolution of an AR using only the LoS magnetic field as the lower boundary condition combined with a potential field as the initial condition, such a simple model does not work in all cases. There were several issues that were encountered in the simulation when using the simplest initial and boundary conditions (an initial potential field condition and closed top boundary). Firstly, the presence of highly twisted field near the side boundaries of the box. Boundary effects can be rectified by re-scaling the magnetograms to occupy a smaller area at the bottom of the computational box during the clean-up procedure (Appendix~\ref{sec:AppA}). If the magnetograms contain large amounts of small-scale magnetic field that affect the simulated coronal evolution, these can be removed by smoothing the magnetograms with a Gaussian kernel (see Appendix~\ref{sec:AppB}). This process is applied in addition to the clean-up procedure detailed in Appendix~\ref{sec:AppA}. If the simulation runs for long time periods, twisted magnetic field can build-up in the computational volume. By adding coronal diffusion, in the form of Ohmic diffusion or hyperdiffusion, this can help prevent the build-up of highly twisted field by decreasing the amount of poloidal flux. However, despite the inclusion of additional coronal diffusion terms, flux ropes are still able to form and reach instability in the simulation and the overall evolution of the simulated coronal field remains significantly unaffected \citep{Mackay-2006a, Yardley-2018b}. 


The energy and non-potentiality of the coronal field in the simplest simulation setup only originates from the Poynting flux due to horizontal motions. For the cases where the simple model is insufficient to describe the observations (ARs with a score of 2 or 3) there could be additional physical effects that are acting. For example, the initial configuration of the coronal magnetic field could be non-potential and therefore a LFF field initial condition could be implemented to represent any non-potential effects present before the start of the simulation. 
When a LFF field initial condition is used the force-free parameter $\alpha$ is assigned a small value with a magnitude of $10^{-9}$--$10^{-8}$~m$^{-1}$ (see Table~\ref{tab:table2}), to match the weak shear seen in the coronal observations. The range in the force-free parameter is constrained by the size of the computational domain which scales as 1/L, where L varies from one AR to the next. This is due to the nature of the LFF field solution requiring a decaying (non-oscillatory) solution with height. The sign of $\alpha$ is taken from the sense of twist from the magnetic tongues present in the observations \citep{Luoni-2011}. The sign and value of $\alpha$ in our simulations is therefore selected in a similar manner to our previous study \citep{Yardley-2018b}.

There may also be other sources of energy or helicity injection, which are not captured by the evolution of the normal component of the magnetic field that have to be taken into account, such as the presence of vertical motions or torsional Alfv{\'e}n waves. Along with these additional injection mechanisms non-ideal processes may also have to be considered. These effects are implemented one at a time in the simulation by modifying the induction equation to include the physical effects through three additional terms: 
\begin{equation}
\frac{\partial \boldsymbol{A}}{\partial t} = \boldsymbol{v} \times \boldsymbol{B} - \eta \boldsymbol{j} + \frac{\boldsymbol{B}}{B^{2}} \nabla \cdot (\eta_{4} B^{2} \nabla \alpha) - \nabla_{z} (\zeta B_{z}), \label{eq:global}
\end{equation}
where
\begin{equation}
\alpha = \frac{\boldsymbol{B} \cdot \nabla \times \boldsymbol{B}}{B^{2}}.
\end{equation}
The first additional term is Ohmic diffusion, where $\eta$ represents the resistive coefficient. The second additional term, is hyperdiffusion \citep{Boozer-1986, Strauss-1988, Bhattacharjee-1995}. This diffusion term is artificial and is introduced to reduce gradients that are present in the force-free parameter $\alpha$, while total magnetic helicity remains conserved \citep{vB-2007}. 

The third additional term represents the injection of a horizontal magnetic field or twist component at the photospheric boundary. In this term $\nabla_{z}$ is the vertical component of the gradient operator and $\zeta$ is an injection parameter that has the dimensions of a diffusivity. The parameter $\zeta$ is only non-zero at the photospheric boundary ($z=0$) hence, the injection of the horizontal field only occurs at this location. This term leads to a change in $A_{z}$ half a grid point into the domain, and the subsequent injection of a horizontal magnetic field and magnetic helicity into the corona. By applying this injection in $A_{z}$ leaves the vertical component of the magnetic field unchanged. The sign of the injection parameter $\zeta$ determines the sign of the magnetic helicity that is injected via the horizontal field. A positive (negative) value of $\zeta$ leads to the injection of negative (positive) magnetic helicity. Once injected, the horizontal field and twist component propagate upwards along the magnetic field lines through the $\mathbf{v} \times \mathbf{B}$ term in the induction equation above (Equation~\ref{eq:global}). This term is mathematically equivalent to that used in \citet{Mackay-2014} to model the helicity condensation process of \citet{Antiochos-2013}. For the present simulations this term does not represent helicity condensation rather it is used to add an additional non-potential contribution that is not captured by a potential field initial condition or the applied horizontal motions on the photospheric surface alone. Additional sources of helicity may originate from the prior evolution of an AR that is not captured from the initial potential field, the presence of vertical motions or the propagation of torsional Alfv{\'e}n waves from below the photosphere into the corona. The additional injection of horizontal magnetic field at the photosphere, along with the Ohmic and hyperdiffusion terms are included in the simulation through user-defined constants.

\subsubsection{Magnetic Field Evolution}
We now modify the top boundary, initial condition and include non-ideal terms in the simulations. This is to determine whether it is possible to improve the simulation results, obtained for the ARs in Section~\ref{sec:simp}, where only a reasonable or minimal amount of the coronal structure was captured (ARs assigned a scoring criteria of 2 or below).

\begin{figure}
\centerline{\includegraphics[width=1\textwidth,clip=]{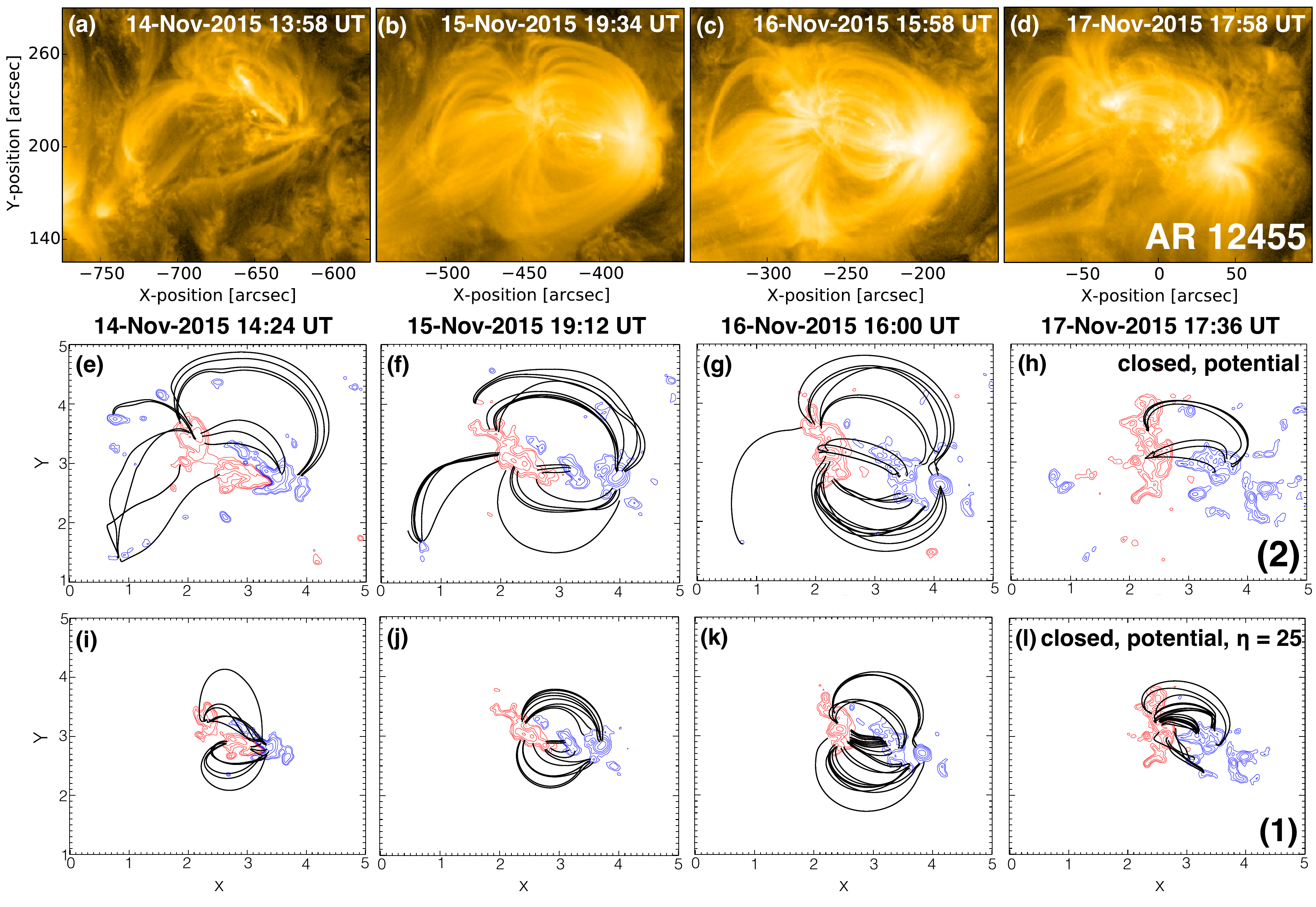}}
\caption{The SDO/AIA 171~\AA\ images (a--d) and corresponding sample field lines (e--l) showing the evolution of AR 12455. The second row (e--h) shows sample field lines from the simulation run with closed top boundary conditions and an initial potential field i.e. the simplest initial and boundary conditions. The third row (i--l) shows the results when Ohmic diffusion, $\eta$ is added with a value of 25~km$^{2}$~s$^{-1}$ and small-scale field has been removed. The score given to each simulation is given at the bottom right of panels (h) and (l).
The positive (negative) photospheric magnetic field is represented by the red (blue) contours. }
\label{fig5}
\end{figure}

To improve the results obtained by using the simplest initial and boundary conditions additional physical effects, Gaussian smoothing, and LFF field initial conditions are used. The simulations that were performed for each AR to improve the previous results are described in the comments (final) column of Table~\ref{tab:table2}. If the performance of the simulation improved, the new score is included in brackets in the Score (fourth) column of Table~\ref{tab:table2}. An example can be seen in Figure~\ref{fig5} where AR 12455 improves from a score of 2 to 1. The original simulation captured the evolution of the large-scale coronal loops in the north of the AR relatively well however, failed to reproduce the large-scale loops present in the south. It also failed to capture the bright core of the AR (see Figure~\ref{fig5} (a)). By removing the small-scale magnetic field at the AR periphery using a Gaussian kernel and then introducing Ohmic diffusion the simulation is able to replicate the small and large-scale coronal structure for the entire AR evolution including the sheared structure present at the start of the AR evolution.

The new results are as follows. The enhanced simulations are able to capture the majority of the coronal structure for 12 ARs, a reasonable amount of structure for five ARs and little or no coronal structure for three ARs (see brackets in Table~\ref{tab:table3}). The new results show that one AR moved from scoring category 3 to 2 and two ARs moved from 2 to 1 indicating there was an overall improvement of 20\% when a mix of additional physical effects are included. Therefore, the NLFF field simulation is able to capture a reasonable amount of the structure for 85\% of the ARs and only failed to capture the structure for 15\% of the ARs from the sample. This is a slight improvement on the previous result, where the simplest initial and boundary conditions were used. The improvement in the results is mainly due to the use of a LFF field initial condition. Although, the application of Gaussian smoothing to remove additional small-scale magnetic field near the AR periphery and the addition of Ohmic diffusion also improved the results. When considering the build-up to eruption in the simulations, no improvement is made on the previous results as the simulations again successfully follows the build-up to eruption for 12 out of the 24 observed eruptions.

\section{Discussion} \label{sec:dis}

We have used the method of \citet{Mackay-2011} to simulate the full coronal evolution of 20 bipolar ARs, from emergence to decay, using a time series of LoS magnetograms as the lower boundary condition. To reproduce the full coronal evolution of the ARs requires a series of magnetograms that extends over the entire lifespan of each AR.

Numerous clean-up processes (see Appendix~\ref{sec:AppA}) have been applied to the raw magnetograms including time-averaging, removal of isolated features, removal of low flux values, and flux balancing before carrying out the simulations. The application of these procedures produces a series of cleaned magnetograms with a smooth and continuous evolution of the photospheric magnetic field. By using a series of cleaned magnetograms as the lower boundary condition it is easier to simulate the large-scale coronal magnetic field evolution of the ARs as the inclusion of small-scale magnetic elements and random noise could potentially lead to numerical problems in the simulations.


The method has not yet been tested using vector magnetograms as the lower boundary conditions of the simulation. However, an initial qualitative comparison between the vector components at the simulation boundary to the observed vector data of one AR (AR 11561) in our sample shows a relatively good agreement (see Appendix~\ref{sec:AppC} for more details). In a follow-up study we will expand on this qualitative comparison between the simulated and observed vector magnetic field components.

Initially, the ARs were simulated using the simplest initial and boundary conditions i.e. a potential field initial condition and closed top boundary. We conclude, after a manual comparison with the observations, that the simulations reproduced a reasonable amount of the coronal structure and evolution for 80\% of the ARs. This result is improved slightly to 85\% by applying Gaussian smoothing to remove additional small-scale magnetic field in the magnetograms, using a LFF field initial condition, and including additional effects such as non-ideal terms in the simulations. For the ARs where the simulation failed to reproduce the main coronal features, particularly during the early stages of the AR evolution, a NLFF field initial condition may be more appropriate. We will implement the use of a NLFF field initial condition in the future by constructing a NLFF field extrapolation using the technique described by \citet{Valori-2005}. Whereby a potential field will be extrapolated from a magnetogram, the horizontal field components will be set using a vector magnetogram, and magnetofrictional relaxation will be applied to relax the magnetic field to a force-free equilibrium.

We do not constrain the simulations with coronal observations therefore, the coronal structures reproduced by the simulation are the result of the non-potential effects produced by the boundary evolution. Therefore, the accuracy of the coronal field models in this study has been judged qualitatively by a manual inspection and visual comparison to the coronal observations. To make the comparison to observations less time-consuming and to remove the subjective nature of this analysis an optimization method could be developed to minimize the deviation between the field lines from the simulation and the intensity observations. An optimization technique will be considered in future studies.

We now discuss when and where the NLFF field simulations were able to reproduce the build-up to eruption. By reproducing the build-up to eruption, we are referring to the ability to identify a flux rope that has formed in the simulation that loses equilibrium or becomes unstable at the same location and at a similar time to the eruption that occurred in the observations. We do not aim to recreate the full dynamics of the eruptions as this requires a MHD simulation.

The simulations were able to replicate the formation and eruption onset of flux rope structures at the internal PIL of an AR where the flux rope was created by flux cancellation and magnetic reconnection occurring at low atmospheric heights. Signatures of eruption onset were found in the simulations for all four low corona eruptions that originated from the internal PIL of ARs 11437, 11561, 11680, and 12382. The simulations were analysed within a $\pm$~12~hr window of the eruption occurring in the observations and the mean unsigned time difference of eruption onset taking place in the simulations compared to the observed eruptions in these four ARs was found to be $\approx$5~hrs. These simulation results support the \citet{vB-1989} scenario and show that the physical processes can be replicated on a similar timescale to that which the Sun evolves over.

The technique failed to capture the onset of some of the eruptions that originated from low in the corona along an external PIL or at high-altitudes. There are a number of possible reasons for this. Capturing the initiation of eruptions that occurred during the early stages of the simulations proved challenging since these eruptions occur during the flux emergence phase of the ARs.
Using an initial potential field condition, combined with the short time over which the coronal field is being evolved, means that insufficient shear and free energy will have built-up in the simulated coronal field.

To combat this issue we can vary the initial or boundary conditions and include additional non-ideal effects in the simulation. For six of the ARs we constructed a LFF field initial condition to see how this affected the results. We chose the magnitude and sign of the force-free parameter to reflect the weak shear seen in the coronal observations. Ideally, vector data can be used to calculate the value of $\alpha$ to use to construct the LFF field initial conditions for the simulation. In the future, we aim to use the observed $\alpha$ value for the LFF field initial condition in our simulations or use a NLFF field initial condition when possible.


The simulation method also fails to capture the eruption onset for ejections that occur in quick succession as it is impossible to separate them from one another in the simulation. To recreate the dynamics of multiple eruptions over short timescales requires the use of full MHD simulations.
For example, four eruptions from external PILs were observed to occur in quick succession during the first 12 hrs of the emergence phase of AR 12229. The build-up to these eruptions was not captured by the simulation and this AR accounted for a large number of the missed eruptions. There was also a large imbalance in the magnetic flux during emergence due to the AR emerging at $\approx$50$^{\circ}$ longitude into negative quiet Sun magnetic field.
Additionally, eruptions that originated along an external PIL were observationally found to occur due to flux cancellation that takes place between the periphery of the AR and quiet Sun magnetic field during the emergence phase \citep{Yardley-2018a}. Eruptions that form at the external PIL are harder to simulate because much of the small-scale field is removed during the ``cleaning" process or is not included in the local simulations. At present the simulation method is designed to capture the local and internal evolution of the ARs. 


In \citet{Yardley-2018a} the origin of each high-altitude event was not studied in detail but it was suggested that they could be the result of the formation of a high-altitude structure during the evolution of the AR or the destabilization of a pre-existing external structure. The build-up to the high-altitude eruptions that were observed in ARs 11446, 11886, and 12336 were not replicated by the simulations. This could be taken into account in future work by using a NLFF field initial condition on a case-by-case basis if a flux rope is present at the start or early stages of the simulation. However, if the high-altitude events are a result of the destabilisation of pre-existing structures then this technique will not be able to capture their formation. Therefore, to be able to capture the onset of eruptions that arise due to the interaction of external magnetic fields or large-scale coronal structures, non-local effects need to be taken into account by using global NLFF field models (e.g. \citealt{Mackay-2006a}) to simulate the evolution of the large-scale corona.

Presently, we have focused on simulating the evolution of a set of relatively simple, bipolar ARs that produce faint eruption signatures and a limited number of CMEs. However, this is necessary to test the method before simulating larger, more complex ARs. In the future we will simulate a broader range of ARs, including multipolar regions and large AR complexes that produce multiple CMEs. Given the results of the applied technique simulating larger, multipolar and non-isolated regions should be possible but will require a larger computational domain.


\section{Summary}\label{sec:sum}

In this study, the coronal evolution of 20 bipolar ARs was simulated from emergence to decay. The simulations were carried out in order to test whether the evolution of the coronal magnetic field through a series of NLFF states driven by boundary motions could successfully reproduce the observed coronal features of the ARs and the onset of eruption. The coronal magnetic field evolution was simulated by applying the NLFF field method of \citet{Mackay-2011} to LoS magnetograms taken by {\it SDO}/HMI that were used as the lower boundary conditions. The simulated coronal field evolution for each AR was manually compared to the 171 and 193~\AA\ emission structures as seen by SDO/AIA. 

The first simulation results were obtained using the simplest initial and boundary conditions i.e. a potential field initial condition and a closed top boundary. By using this approach it was possible to reproduce a reasonable amount of the coronal structure and evolution for 80\% of the AR sample. In total, the build-up to eruption was successfully followed in the simulations within a $\pm$~12~hr window of the eruptions occurring in the observations for 12 out of the 24 (50\%) of the observed eruptions. 

To improve the simulation results we varied the boundary (from closed to open) and initial condition (from potential to LFF) and included additional parameters such as Ohmic diffusion, hyperdiffusion, and an additional injection of horizontal magnetic field and magnetic helicity in the simulations. We also took into account boundary effects by re-scaling the magnetogram at the bottom of the computational box and removed small-scale magnetic features that affect the large-scale evolution of the coronal field by applying Gaussian smoothing to the magnetograms. These steps were in addition to the clean-up procedures and were carried out one at a time. Through considering various combinations of additional terms there was a slight improvement in the results, as one AR moved from scoring category 3 to 2 and two ARs moved from category 2 to 1. Therefore, by varying the boundary and initial conditions and including additional physical effects in the simulation there was an overall improvement of 20\%. Overall, the simulations were able to capture a reasonable amount of coronal structure for 85\% of the AR sample, only failing to capture the structure of 15\% of the regions. Despite varying the boundary and initial conditions and including additional global parameters the simulations are only able to successfully follow the build-up to eruption for 50\% of the observed eruptions associated with the AR sample. For the successful cases, the key component in reproducing the coronal evolution and build-up to eruption for the ARs is the use of LoS magnetograms, as the lower boundary conditions to the simulations as changing the side/top boundary conditions, initial condition and including additional physical affects had an insignificant effect on the simulated coronal field evolution.

The unsigned mean time difference between the signatures of eruption onset in the simulations compared to the observed eruptions was $\approx$5~hrs. The simulations were carried out over a time period of roughly 96--120 hrs therefore, a mean time difference between eruption onset occurring in the observations compared to the simulations of $\approx$5~hrs is a very favourable result (within 3 applied magnetograms). As current space weather forecasting methods can only provide a warning post-eruption and 1--3 days before the arrival of a CME at Earth with an uncertainty of 12 hrs, our results are well within the present time error. Also, as our approach is computationally efficient we can reproduce the coronal magnetic field evolution of ARs over several days within a few hours of computation time on a desktop machine. 

In fact, \citet{Pagano-2019a,Pagano-2019b} have demonstrated how eruption metrics based on the NLFF field simulations may be used to distinguish eruptive from non-eruptive ARs. This work has also demonstrated how it is possible to provide near-real time alerts of eruptions using the observed LoS magnetograms, the NLFF field simulations and the projection of the simulations forward in time. The analysis carried out in these studies includes four ARs taken from our AR sample in this paper. The initial results from \citet{Pagano-2019a,Pagano-2019b} are promising but additional work is required, including addressing the issues outlined in Section~\ref{sec:dis}, before the method can identify the exact eruption time and be implemented for CME forecasting purposes.


In summary, the full coronal magnetic field evolution of 20 bipolar ARs was simulated using the time-dependent NLFF field method of \citet{Mackay-2011}. Using this method, it was possible to reproduce the main coronal features present in the observations for 85\% of the AR sample. The simulations were also able to successfully follow the build-up to and onset of eruption within a $\pm$12~hr window for 12 out of the 24 eruptions (50\%) that were identified in the observations. The mean unsigned time difference between the eruptions occurring in the observations compared to the time of eruption onset in the simulations was found to be $\approx$5~hrs. It is important to acknowledge that for all four eruptions that took place along the internal PIL of the ARs, the simulations were able to model the timings of eruption onset with a mean unsigned time difference of $\approx$7~hrs. Therefore, the simulations were able to successfully reproduce the local evolution for the majority of the ARs in the sample.





\begin{acks}
The authors would like to thank SDO/HMI and AIA consortia for the data, and also for being able to browse this data through JHelioviewer (http://jhelioviewer.org, \citealt{Muller-2017}). The analysis in this paper has made use of SunPy, which is an open-source and free community-developed solar data analysis package written in Python \citep{SunPy-2015}. S.L.Y. would like to acknowledge STFC for support via the Consolidated Grant SMC1/YST025. D.H.M. would like to thank STFC, the Leverhulme Trust and the ERC under the Synergy Grant: The Whole Sun, grant agreement no. 810218 for financial support. L.M.G. is thankful to the Royal Society for a University Research Fellowship and the Leverhulme Trust.
\end{acks}

{\bf Disclosure of Potential Conflicts of Interest} The authors declare that they have no conflicts of interest.


\begin{landscape}
\begin{footnotesize}
\setlength\LTcapwidth{\linewidth}
\begin{longtable}{*{9}{c}}


\caption{The 20 bipolar ARs simulated in this study. The table includes the NOAA number assigned to the AR and the heliographic coordinates of the AR at the time of emergence. The value of peak unsigned flux (half the total absolute positive and negative flux), the start of emergence, peak unsigned flux and end of observation times are also given. The timings of the events that originate from low altitude along the internal PIL, along external PILs, and from high altitude are listed in the final columns. The time and GOES class of four flares and the timings of the two CMEs that are observed in LASCO/C2 that are associated with the ARs are given in the footnotes.
The AR properties in this Table have been taken from the previous study of \citet{Yardley-2018a}.}
\label{tab:table1}\\

\hline
NOAA & Hel. & Magnetic & Emergence Start & Peak Flux & End & Internal PIL & External PIL & High Altitude \\
AR & Coords. & Flux & Time & Time & Time & Event Timings & Event Timings & Event Timings  \\
No. & ($\theta$,$\phi$) & (10$^{21}$~Mx) & (UT) & (UT) & (UT) & (UT) & (UT) & (UT) \\
\hline
\endfirsthead
11437 & S29 E33 & 0.6 & 16-Mar-12 12:46 & 17-Mar-12 15:58 & 21-Mar-12 01:34 & 20-Mar-12 14:46 & 17-Mar-12 05:14 & - \\
& & & & & & 17-Mar-12 10:53 & - \\ 
11446\footnote{GOES B4.0 class flare, 26-Mar-12 13:17 UT} & N31 E20 & 1.0 & 22-Mar-12 15:58 & 24-Mar-12 15:58 & 26-Mar-12 23:58 & - & - & 24-Mar-12 00:42 \\
11480 & S14 E26 & 0.8 & 09-May-12 11:10 & 11-May-12 04:46 & 13-May-12 23:58 & - & - & - \\
11561 & S18 E34 & 1.5 & 29-Aug-12 19:10 & 31-Aug-12 04:46 & 01-Sep-12 23:58 & 01-Sep-12 23:37 & - & - \\
11680 & S25 E52 & 2.2 & 24-Feb-13 14:22 & 25-Feb-13 14:22 & 03-Mar-13 17:34 & 03-Mar-13 17:27 & - & - \\
11808\footnote{GOES B8.7 class flare, 30-Jul-13 20:49 UT} & N12 E66 & 3.9 & 29-Jul-13 01:34 & 30-Jul-13 01:34 & 07-Aug-13 23:58 & - & - & 30-Jul-13 04:04 \\
& & & & & & & & 31-Jul-13 15:10 \\
11813\footnote{GOES B5.5 class flare, 08-Aug-13 16:14 UT} & S19 E22 & 2.7 & 06-Aug-13 01:34 & 08-Aug-13 17:34 & 11-Aug-13 23:58 & - & - & - \\
11867 & N17 E05 & 3.3 & 11-Oct-13 07:58 & 13-Oct-13 15:58 & 14-Oct-13 04:46 & - & - & - \\
11881 & S25 E52 & 1.4 & 24-Oct-13 01:34 & 26-Oct-13 15:58 & 29-Oct-13 23:58 & - & - & 24-Oct-13 08:10\footnote{LASCO/C2 CME, 24-Oct-13 09:12 UT} \\
& & & & & & - & - & 27-Oct-13 19:45 \\
& & & & & & - & - & 29-Oct-13 02:58 \\
11886 & N10 E14 & 2.3 & 28-Oct-13 09:34 & 30-Oct-13 17:34 & 01-Nov-13 22:22 & - & - & 29-Oct-13 12:57 \\
12086 & N03 E49 & 0.9 & 08-Jun-14 15:58 & 09-Jun-14 23:58 & 11-Jun-14 23:58 & - & 10-Jun-14 14:49 & - \\
12119 & S26 E38 & 2.5 & 18-Jul-14 04:46 & 21-Jul-14 06:22 & 23-Jul-14 23:58 & - & 18-Jul-14 10:40 & -  \\
& & & & & & - & 22-Jul-14 21:02 & -\\
& & & & & & - & 23-Jul-14 07:30\footnote{LASCO/C2 CME, 23-Jul-14 08:12 UT} & - \\
12168 & N10 E08 & 2.2 & 16-Sep-14 12:46 & 18-Sep-14 11:10 & 23-Sep-14 23:58 & - & - & - \\
12229 & S23 E50 & 0.8 & 04-Dec-14 20:46 & 5-Dec-14 22:22 & 09-Dec-14 12:46 & - & 05-Dec-14 03:46 & - \\
& & & & & & - & 05-Dec-14 08:12 & - \\
& & & & & & - & 05-Dec-14 10:39 & - \\
& & & & & & - & 05-Dec-14 12:35 & - \\
12273\footnote{GOES C2.4 class flare, 27-Jan-15 05:43 UT} & N02 E21 & 2.7 & 25-Jan-15 15:58 & 27-Jan-15 19:10 & 31-Jan-15 01:34 & - & - & - \\
12274 & N03 E09 & 0.4 & 25-Jan-15 17:34 & 26-Jan-15 15:58 & 28-Jan-15 17:34 & - & 25-Jan-15 20:00 & - \\
12336 & N17 E49 & 2.9 & 01-May-15 14:22 & 05-May-15 20:46 & 08-May-15 23:58 & - & - & 05-May-15 01:29 \\
& & & & & & - & - & 05-May-15 09:24 \\
12382 & S08 E29 & 0.6 & 04-Jul-15 03:10 & 05-Jul-15 20:46 & 09-Jul-15 12:46 & 09-Jul-15 02:29 & - & - \\
12453 & N04 E29 & 1.7 & 12-Nov-15 07:58 & 15-Nov-15 15:58 & 16-Nov-15 23:58 & - & - & - \\
12455 & N14 E61 & 1.4 & 13-Nov-15 04:46 & 16-Nov-15 03:10 & 18-Nov-15 23:58 & - & - & - \\
\hline
\end{longtable}
\end{footnotesize}
\end{landscape}

\begin{landscape}



\setlength\LTcapwidth{\linewidth}
\begin{longtable}{>{\centering}p{1.2cm}>{\centering}p{1.6cm}>{\centering}p{1.6cm}>{\centering}p{1cm}>{\centering}p{1.6cm}>{\centering}p{1.6cm}p{8cm}}

\caption{Results of the NLFF field simulations. The table includes the NOAA AR number, the number of time steps or magnetograms used to simulate the coronal evolution, and whether there was a flux imbalance present between the positive and negative photospheric polarities of the AR. This is followed by the agreement between the simulations and observations, the number of observed eruptions the simulation can follow the build-up to eruption for, and the time difference between the signatures of flux rope eruption onset that occurred in the simulations and the eruptions in the observations. The final column gives additional information such as the location of the AR during emergence if close to 60$^{\circ}$, as well as the surrounding magnetic field environment the region emerges into. It also gives the initial conditions, boundary conditions and additional global parameters that were used to improve the performance of the simulation. If improvements were made, the new score is given in brackets in column four.} 
\label{tab:table2}\\
\hline
NOAA & No. of & Flux  & & No. of  & Time & \\
AR & timesteps & Imbalance & Score & Eruptions & Diff. (hrs) & Comments \\
\hline
\endfirsthead

11437 & 62 & small & 1 & 2/3 & -1, -10 & - \\
11446 & 62 & large & 2 & 0/1 & - & Region dominated by positive quiet Sun magnetic field. Re-scaled boundary condition. LFF field initial condition is used with $\alpha = 1.18 \times 10^{-8}$~m$^{-1}$. \\
11480 & 69 & small & 2 & - & - & Re-scaled boundary condition. Gaussian smoothing applied to remove small-scale field. \\
11561 & 50 & small & 1 & 1/1 & -4 & - \\
11680 & 109 & small & 1 & 1/1 & -9 & - \\
11808 & 60 (48) & small & 3(2) & 1/2 & 12 & Emerges before 60$^{\circ}$, started simulation at a later time. LFF field initial condition is used with $\alpha = 7.98 \times 10^{-9}$~m$^{-1}$. \\
11813 & 90 & small & 1 & - & - & Re-scaled boundary condition. LFF field initial condition is used with $\alpha = -6.42 \times 10^{-9}$~m$^{-1}$. Gaussian smoothing applied to remove small-scale field.\\
11867 & 43 & large & 2(1) & - & - & Emerges into positive quiet sun magnetic field. LFF field initial condition used with $\alpha = -5.64 \times 10^{-9}$~m$^{-1}$. Gaussian smoothing applied to remove small-scale field.\\
11881 & 86 & small & 1 & 2/3 & 2, 1 & - \\
11886 & 69 & small & 2 & 0/1 & - & Emerges into negative quiet Sun magnetic field. LFF field condition is used with $\alpha = -8.91 \times 10^{-9}$~m$^{-1}$.\\
12086 & 66 (51) & small & 3 & 0/1 & - & Emerged into negative quiet Sun magnetic field. Re-scaled boundary condition. Stopped simulation at an earlier time step due to emergence. \\
12119 & 85 & small & 1 & 3/3 & 6, -1, 0 & - \\
12168 & 107 & large & 1 & - & - & Emerges into negative quiet Sun magnetic field. \\
12229 & 71 & large & 3 & 0/4 & - & Emerged into negative quiet Sun magnetic field. Hard to compare to observations as limited coronal loops are visible. Gaussian smoothing applied to remove small-scale field.\\
12273 & 79 & small & 3 & - & - & Large negative sunspot, with highly twisted structure along the internal PIL. Ohmic diffusion was applied where $\eta =$~25~km$^{2}$~s$^{-1}$ and Gaussian smoothing was applied to remove small-scale field. An additional injection of magnetic helicity was applied where $\zeta =$[-1,-10]~km$^{2}$~s$^{-1}$. \\
12274 & 46 & large & 1 & 1/1 & 5 & Emerged into negative quiet Sun magnetic field. Hard to compare to observations as limited coronal loops are visible. \\
12336 & 111 & small & 2 & 0/2 & - & LFF field initial condition used with $\alpha= -8.89 \times 10^{-9}$~m$^{-1}$. Gaussian smoothing applied to remove small-scale field. \\
12382 & 78 & small & 1 & 1/1 & -5 & - \\
12453 & 70 & large & 1 & - & - & Emerged into negative quiet Sun magnetic field. 
Very twisted structure present at the end of the evolution due to small-scale magnetic field. Gaussian smoothing applied to remove small-scale field.\\
12455 & 86 & small & 2(1) & - & - & Emerged close to 60$^{\circ}$. Very twisted structure present at the end of evolution. Re-scaled and coronal diffusion added.\\
\hline
\end{longtable}
\end{landscape}

\begin{table}
\caption{The simulation performance results. The number and percentage of ARs in each scoring category when using the simplest initial and boundary conditions in the simulation are given. The results for the simulations where additional global parameters, Gaussian smoothing and LFF field conditions are used are given in brackets. }
\label{tab:table3}
\begin{tabular}{cccc}
\hline
Score & 1 & 2 & 3 \\
\hline
No. of ARs & 10 (12) & 6 (5) & 4 (3) \\
Percentage (\%) & 50 (60) & 30 (25) & 20 (15) \\
\hline
\end{tabular}
\end{table}

\appendix

\section{Clean-up Processes} \label{sec:AppA}

To analyse the coronal evolution of the 20 bipolar ARs magnetogram data taken from the SDO/HMI instrument were utilised. Full-disk LoS magnetograms were used from the 720~s data series (hmi.M\_720s), which have a pixel size and a noise level of 0.5" and 10~G, respectively. The number of magnetograms used to study the evolution of each AR varied depending upon the ARs lifetime and selection criteria. We apply the following cosine correction before the clean-up procedures are implemented to estimate the radial magnetic field component:
\begin{equation}
    B_{R} = \frac{B_{\rm LOS}}{\cos{\theta} \cos{\phi}},
\end{equation}
where $B_{\rm LOS}$ is the line-of-sight magnetic field and $\theta$ and $\phi$ are expressed in heliocentric coordinates (see Section 2.3 of \citet{Yardley-2018a} for further details). Each corrected magnetogram is then differentially rotated to account for area foreshortening that occurs at large distances from central meridian. Cut-outs of the corrected, de-rotated magnetograms are taken, centred on the AR. This is a relatively simple correction however, this method has been used in previous studies (for example see \citealt{Yardley-2018b}).

\begin{figure}
\centerline{\includegraphics[width=1\textwidth,clip=]{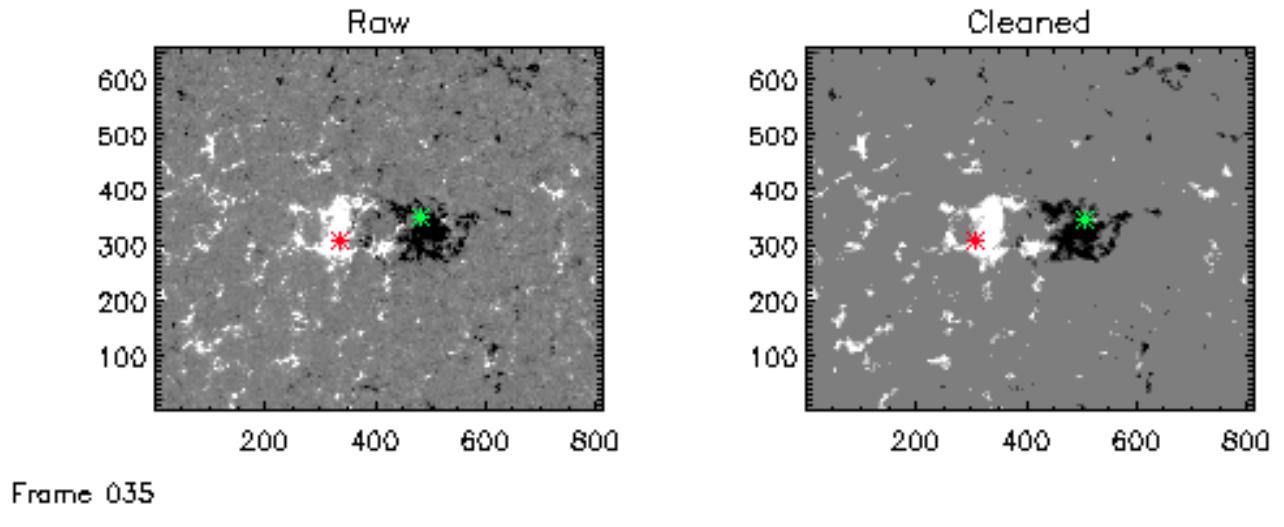}}
\caption{The raw and cleaned magnetograms for AR 11867 taken at 15:59~UT on 2013 October 13 when the region reaches its maximum unsigned magnetic flux. The saturation level of both the magnetograms is $\pm$100~G. The flux-weighted central coordinates for the positive and negative photospheric magnetic polarities are represented by the red and green asterisks, respectively.}
\label{figA1}
\end{figure}

The clean-up procedure described in this section is comparable to the procedure that is used in both \citet{Gibb-2014} and \citet{Yardley-2018b}. The noise level in the raw magnetograms is high, particularly in the early and late stages of AR evolution when the AR is located far from central meridian. Therefore, before the magnetograms are used as the lower boundary conditions of the simulation a number of clean-up processes are implemented. First, the magnetograms are time-averaged by applying the following Gaussian kernel

\begin{equation}
C_{i} = \frac{ \sum^{n}_{j=1}  \mathrm{exp}(-[i-j]/\tau)^{2} F_{j} } {\sum^{n}_{j=1} \mathrm{exp}(-[i-j]/\tau)^{2} },
\end{equation}

where $C_{i}$ is the $i$th cleaned frame and takes values between 1 to $n$ where $n$ is the number of magnetograms in the sequence. $F_{J}$ is the $j$th raw frame, and $\tau$ represents the frame separation where the weighting decreases by $ 1/ \mathrm{e}$. In this study, the frame separation is set to two meaning that each cleaned frame is a linear combination of the total number of raw frames where the two frames before and after the current frame are weighted the highest. This procedure removes random noise and retains the large-scale features of the ARs. As previously stated, this study focuses on the large-scale evolution of the AR magnetic field and not small-scale elements of the quiet Sun.

The next step in the clean-up process includes the removal of small-scale isolated field pixel-by-pixel by evaluating the eight nearest neighbours of each pixel. When fewer than four of the neighbouring pixels have the same sign of magnetic flux then the value of magnetic flux of that pixel is set to zero. Therefore, the pixels at the edge of the magnetogram also have their values set to zero as they have less than four nearest neighbours. In addition, any pixels that have a magnetic flux value below a 25~Mx~cm$^{-2}$ threshold are part of the background magnetic field of the quiet Sun and are also set to zero. At this point the user can choose how to place the magnetograms within the box i.e. the magnetograms can be scaled up/down to fit the computational box or a custom scaling can be applied. In this study, to avoid boundary effects, we rescale the magnetograms to fill 60--70\% of the computational box.

The last clean-up process is implemented when the top boundary condition in the simulation is set to closed and the magnetograms need to be flux balanced. To flux balance the magnetograms the signed magnetic flux of each frame is calculated. The pixels of non-zero value are summed for each frame and the signed magnetic flux is divided by this total. From every pixel that has a non-zero value the imbalanced magnetic flux per pixel is deducted. As the maximum correction is less than 25~Mx~cm$^{-2}$ no pixels change sign during the balancing of magnetic flux. This is the same threshold that is used to set pixels that form part of the background quiet Sun magnetic field to zero.


\section{Gaussian Smoothing} \label{sec:AppB}

\begin{figure}
\centerline{\includegraphics[width=1\textwidth,clip=]{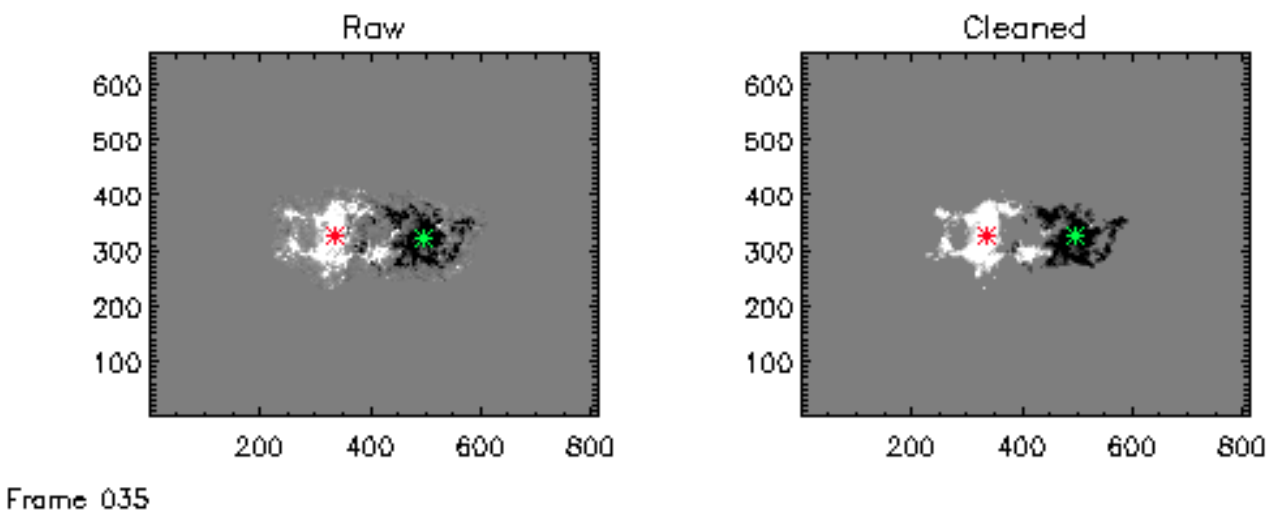}}
\caption{The raw and cleaned magnetograms for simulation frame 35 taken on 2013 October 13 when AR 11867 is at its maximum unsigned magnetic flux value. A Gaussian kernel is used to remove small-scale magnetic field surrounding the AR before the clean-up processes described in Section~\ref{sec:AppA} are applied. For both magnetograms the saturation levels of the photospheric magnetic field are $\pm$100~G. The flux-weighted central coordinates for the positive and negative photospheric magnetic polarities are represented by the red and green asterisks, respectively.}
\label{fig:figB}
\end{figure}

In some cases the small-scale magnetic field at the periphery of the AR is removed before the clean-up procedure is applied (Figure~\ref{fig:figB}). This is achieved by using a method similar to \citet{Yardley-2016}. Firstly, a Gaussian filter is applied with a standard deviation (width) of 7 pixel units to smooth the data. Secondly, the weighted average of the value of magnetic flux density of the neighbouring pixels must exceed a 40~G cut-off. Then, the largest regions that are identified that make up at least 60\% of the selected area are kept whereas, smaller features at large distances from the AR are discarded. This procedure removes small-scale quiet Sun features that are not part of the AR, and does not affect the coronal evolution as we are only interested in the large-scale coronal evolution of the AR.

\section{Comparison to Vector Magnetic Field Observations} \label{sec:AppC}

The method used to simulate the coronal evolution of our AR sample uses a time series of LoS magnetograms as the photospheric boundary condition. This boundary condition injects electric currents into the coronal magnetic field which then evolves through a time series of NLFF fields using the magnetofrictional relaxation process. At no point in the simulation do we constrain the solution using the observed vector magnetic field or with coronal observations. Therefore we allow the boundary evolution to self-consistently produce the horizontal field and subsequently the coronal structures. The reproduced coronal structures are therefore due to non-potential effects produced by the horizontal evolution of the LoS magnetic fields along with any flux emergence or cancellation.
 
\begin{figure*}[h!]
\centerline{\includegraphics[width=0.9\textwidth,clip=]{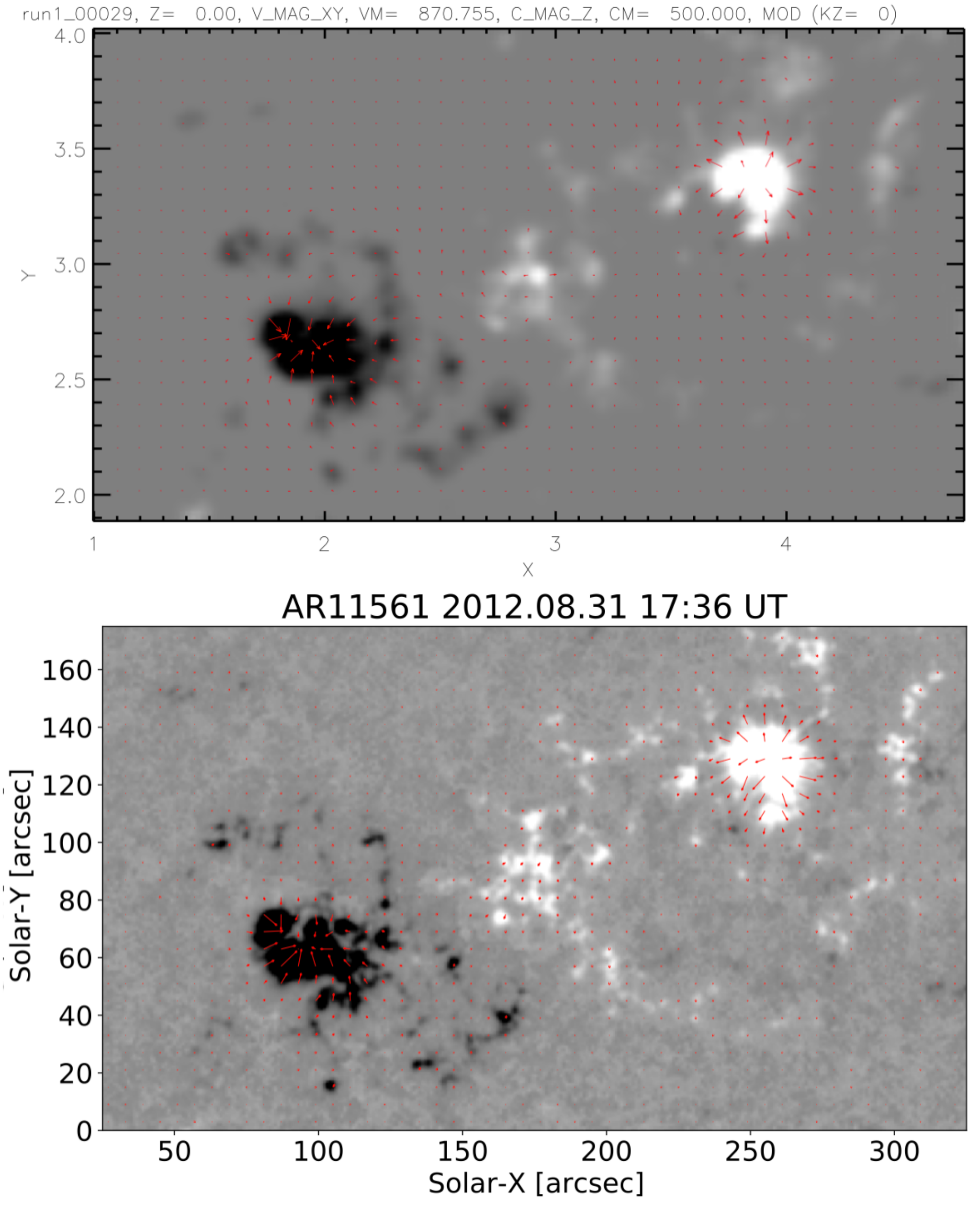}}
\caption{The vertical and horizontal components of the magnetic field of AR 11561 taken from the simulation (top panel) and the corresponding observed vector data (bottom panel). The vertical component of the magnetic field is shown where the positive (negative) photospheric polarities of the AR are represented by the black (white) contours saturated at 500~G. The red arrows represent the magnitude and direction of the horizontal magnetic field components. }
\label{fig:figC}
\end{figure*}

To show that our magnetic field at the photosphere is consistent with the observations and that the simulated coronal structures can be compared with the observed ones we have included a comparison of our simulated vector field at the boundary with the observed vector field. Figure~\ref{fig:figC} shows the vector magnetic field components from the simulation on the base compared to the observed vector data for AR 11561 where the comparison is carried out midway through its evolution. To produce the observed vector field components we have used the space weather HMI active region patches (SHARPS, \citealt{Bobra-2014}) data that have been projected to the Lambert cylindrical equal-area (CEA) Cartesian coordinate system i.e. the hmi.sharp\_cea\_720s series. The figure shows that there is a relatively good agreement between the simulated horizontal field components and the observed horizontal field, particularly in the strong field regions where the signal-to-noise ratio is high. To determine quantitatively whether there is a close correspondence between the horizontal components derived from the base of the simulation and the observed vector components we will conduct an in-depth comparison in a follow-up study. This will include a detailed comparison of the sign and distribution of the three magnetic field components for both the vector field of the simulation and observed vector data.

\end{article} 


\begin{thebibliography}{}

\bibitem[Antiochos(2013)]{Antiochos-2013} Antiochos, S.K.: 2013, {\it \apj} {\bf 772}, 72. \hyperlink{http://doi.org/10.1088/0004-637X/772/1/72}{DOI}.

\bibitem[Arge \emph{et al.}(2004)]{Arge-2004} Arge, C.N., Luhmann, J.G., Odstrcil, D., Schrijver, C.J., and Li, Y.: 2004, {\it \jastp} {\bf 66}, 1295. \hyperlink{http://doi.org/10.1016/j.jastp.2004.03.018}{DOI}.

\bibitem[Arge and Pizzo(2000)]{Arge-2000} Arge, C.N., and Pizzo, V.J.: 2000, {\it jgr} {\bf 105}, 10465. \hyperlink{http://doi.org/10.1029/1999JA000262}{DOI}.


\bibitem[Bhattacharjee and Yuan(1995)]{Bhattacharjee-1995} Bhattacharjee, A., and Yuan, Y.: 1995, {\it \apj} {\bf 449}, 739.
\hyperlink{http://doi.org/10.1086/176094}{DOI}.

\bibitem[Bobra \emph{et al.}(2014)]{Bobra-2014} Bobra, M.G., Sun, X., Hoeksema, J.T., Turmon, M., Liu, Y., Hayashi, K., Barnes, G., and Leka, K.D.: 2014, {\it Solar Physics} {\bf 289}, 3549. \hyperlink{http://doi.org/10.1007/s11207-014-0529-3}{DOI}.

\bibitem[Bobra, van Ballegooijen and DeLuca(2008)]{Bobra-2008} Bobra, M.G., van Ballegooijen, A.A., and DeLuca, E.E.: 2008, {\it \apj} {\bf 672}, 1209. \hyperlink{http://doi.org/10.1086/523927}{DOI}.


\bibitem[Boozer(1986)]{Boozer-1986} Boozer, A.H.: 1986, {\it Journal of Plasma Physics} {\bf 35}, 133. \hyperlink{http://doi.org/
    10.1017/S0022377800011181}{DOI}.

\bibitem[Canou and Amari(2010)]{Canou-2010}Canou, A., and Amari, T.: 2010, {\it \apj} {\bf 715}, 1566. \hyperlink{http://doi.org/10.1088/0004-637X/715/2/1566}{DOI}.

\bibitem[Charbonneau(2010)]{Charbonneau-2010}Charbonneau, P.: 2010, {\it Living Reviews in Solar Physics} {\bf 7}, 3. \hyperlink{http://doi.org/10.12942/lrsp-2010-3}{DOI}. 

\bibitem[Charbonneau(2014)]{Charbonneau-2014} Charbonneau, P.: 2014, {\it Annual Review of Astronomy and Astrophysics} {\bf 52}, 251.
\hyperlink{http://doi.org/10.1146/annurev-astro-081913-040012}{DOI}.

\bibitem[Couvidat \emph{et al.}(2016)]{Couvidat-2016} Couvidat, S., Schou, J., Hoeksema, J.T., Bogart, R.S., Bush, R.I., Duvall, T.L., Liu, Y., Norton, A.A., and Scherrer, P.H.: 2016, {\it \solphys} {\bf 291}, 1887. \hyperlink{http://doi.org/10.1007/s11207-016-0957-3}{DOI}. 

\bibitem[De Rosa \emph{et al.}(2009)]{DeRosa-2009}De Rosa, M.L., Schrijver, C.J., Barnes, G., Leka, K.D., Lites, B.W., Aschwanden, M.J., Amari, T., Canou, A., McTiernan, J.M., R{\'e}gnier, S., Thalmann, J.K., Valori, G., Wheatland, M.S., Wiegelmann, T., Cheung, M.C.~M., Conlon, P.A., Fuhrmann, M., Inhester, B. and Tadesse, T.: 2009, {\it The Astrophysical Journal}, {\bf 696}, 1780. \hyperlink{http://doi.org/10.1088/0004-637X/696/2/1780}{DOI}.

\bibitem[Eastwood \emph{et al.}(2017)]{Eastwood-2017} Eastwood, J. P., Biffis, E., Hapgood, M. A., Green, L. M., Bisi, M. M., Bentley, R. D., Wicks, R., McKinnel, L.-A., Gibbs, M., and Burnett, C.: 2017, {\it Risk Analysis}, {\bf 37}, 206. \hyperlink{http://doi.org/10.1111/risa.12765}{DOI}.

\bibitem[Finn, Guzdar and Usikov(1994)]{Finn-1994} Finn, J.M., Guzdar, P.N., and Usikov, D.: 1994, {\it \apj} {\bf 427}, 475. \hyperlink{http://doi.org/10.1086/174158}{DOI}.

\bibitem[Forbes(2000)]{Forbes-2000} Forbes, T.G.: 2000, {\it \jgr} {\bf 105}, 23153. \hyperlink{http://doi.org/10.1029/2000JA000005}{DOI}.

\bibitem[Forbes and Isenberg(1991)]{Forbes-1991} Forbes, T.G., and Isenberg, P.A.: 1991, {\it \apj} {\bf 373}, 294.
\hyperlink{http://doi.org/10.1086/170051}{DOI}.

\bibitem[Gosling(1993)]{Gosling-1993} Gosling, J.T.: 1993, {\it \jgr} {\bf 98}, 18937.
\hyperlink{http://doi.org/10.1029/93JA01896}{DOI}.

\bibitem[Gibb(2015)]{Gibb-2015} Gibb, G.P.S.: 2015, The Formation and Eruption of Magnetic Flux Ropes in Solar and Stellar Coronae, PhD Thesis.\hyperlink{http://hdl.handle.net/10023/7069}{DOI}.

\bibitem[Gibb \emph{et al.}(2014)]{Gibb-2014} Gibb, G.P.S., Mackay, D.H., Green, L.M., and Meyer, K.A.: 2014, {\it \apj} {\bf 782}, 71. \hyperlink{http://doi.org/10.1088/0004-637X/782/2/71}{DOI}.


\bibitem[Green \emph{et al.}(2018)]{Green-2018} Green, L.M., T{\"o}r{\"o}k, T., Vr{\v{s}}nak, B., Manchester, W., and Veronig, A.: 2018, {\it \ssr} {\bf 214}, 46. \hyperlink{http://doi.org/10.1007/s11214-017-0462-5}{DOI}.


\bibitem[Jiang \emph{et al.}(2014)]{Jiang-2014} Jiang, C., Wu, S.T., Feng, X., and Hu, Q.: 2014, {\it \apjl} {\bf 786}, L16. \hyperlink{http://doi.org/10.1088/2041-8205/786/2/L16}{DOI}.

\bibitem[Kliem \emph{et al.}(2014)]{Kliem-2014} Kliem, B., Lin, J., Forbes, T.G., Priest, E.R., and T{\"o}r{\"o}k, T.: 2014, {\it \apj} {\bf 789}, 46. \hyperlink{http://doi.org/10.1088/0004-637X/789/1/46}{DOI}.

\bibitem[Kliem and T{\"o}r{\"o}k(2006)]{Kliem-2006} Kliem, B., and T{\"o}r{\"o}k, T.: 2006, {\it Phys. Rev. Lett.} {\bf 96}, 255002.
\hyperlink{http://doi.org/10.1103/PhysRevLett.96.255002}{DOI}.

\bibitem[Leka \emph{et al.}(1996)]{Leka-1996} Leka, K.D., Canfield, R.C., McClymont, A.N., and van Driel-Gesztelyi, L.: 1996, {\it \apj} {\bf 462}, 547. \hyperlink{http://doi.org/10.1086/177171}{DOI}.

\bibitem[Lemen \emph{et al.}(2012)]{Lemen-2012} Lemen, J.R., Title, A.M., Akin, D.J., Boerner, P.F., Chou, C., Drake, J.F., Duncan, D.W., Edwards, C.G., Friedlaender, F.M., Heyman, G.F., Hurlburt, N.E., Katz, N.L., Kushner, G.D., Levay, M., Lindgren, R.W., Mathur, D.P., McFeaters, E.L., Mitchell, S. Rehse, R.A., Schrijver, C.J., Springer, L. A., Stern, R.A., Tarbell, T.D., Wuelser, J., Wolfson, C.J., Yanari, C. Bookbinder, J.A., Cheimets, P.N., Caldwell, D., Deluca, E.E., Gates, R., Golub, L., Park, S., Podgorski, W.A., Bush, R.I., Scherrer, P.H., Gummin, M. A., Smith, P., Auker, G., Jerram, P., Pool, P., Soufli, R., Windt, D.L., Beardsley, S., Clapp, M., Lang, J., and Waltham, N.: 2012, {\it \solphys} {\bf 275}, 17. \hyperlink{http://doi.org/10.1007/s11207-011-9776-8}{DOI}.

\bibitem[Longbottom(1998)]{Longbottom-1998} Longbottom, A.: 1998, {\it IAU Colloq. 167: New Perspectives on Solar Prominences}, 274.

\bibitem[Luoni \emph{et al.}(2011)]{Luoni-2011} Luoni, M.L., D{\'e}moulin, P., Mandrini, C.H., and van Driel-Gesztelyi, L.: 2011, {\it \solphys} {\bf 270}, 45. \hyperlink{http://doi.org/10.1007/s11207-011-9731-8}{DOI}.

\bibitem[Mackay, DeVore and Antiochos(2014)]{Mackay-2014} Mackay, D.H., DeVore, C.R., and Antiochos, S.K.: 2014, {\it \apj} {\bf 784}, 164. \hyperlink{http://doi.org/10.1088/0004-637X/784/2/164}{DOI}.

\bibitem[Mackay, Green and van Ballegooijen(2011)]{Mackay-2011} Mackay, D.H., Green, L.M., and van Ballegooijen, A.: 2011, {\it \apj} {\bf 729}, 97. \hyperlink{http://doi.org/10.1088/0004-637X/729/2/97}{DOI}.

\bibitem[Mackay and van Ballegooijen(2006a)]{Mackay-2006a} Mackay, D.H., and van Ballegooijen, A.A.: 2006a, {\it \apj} {\bf 641}, 577.
\hyperlink{http://doi.org/10.1086/500425}{DOI}.

\bibitem[Mackay and van Ballegooijen(2006b)]{Mackay-2006b} Mackay, D.H., and van Ballegooijen, A.A.: 2006b, {\it \apj} {\bf 642}, 1193.
\hyperlink{http://doi.org/10.1086/501043}{DOI}.

\bibitem[Mackay and van Ballegooijen(2009)]{Mackay-2009} Mackay, D.H., and van Ballegooijen, A.A.: 2009, {\it \solphys} {\bf 260}, 321. \hyperlink{http://doi.org/10.1007/s11207-009-9468-9}{DOI}.

\bibitem[Mays \emph{et al.}(2015)]{Mays-2015} Mays, M.L., Taktakishvili, A., Pulkkinen, A., MacNeice, P.J., Rast{\"a}tter, L., Odstrcil, D., Jian, L.K., Richardson, I.G., LaSota, J.A., Zheng, Y., and Kuznetsova, M.M.: 2015, {\it \solphys} {\bf 290}, 1775.
\hyperlink{http://doi.org/10.1007/s11207-015-0692-1}{DOI}. 


\bibitem[Moore \emph{et al.}(2001)]{Moore-2001} Moore, R.L., Sterling, A.C., Hudson, H.S., and Lemen, J.R.: 2001, {\it \apj} {\bf 552}, 833. \hyperlink{http://doi.org/10.1086/320559}{DOI}.

\bibitem[M{\"u}ller \emph{et al.}(2017)]{Muller-2017} M{\"u}ller, D., Nicula, B., Felix, S., Verstringe, F., Bourgoignie, B., Csillaghy, A., Berghmans, D., Jiggens, P., Garc{\'\i}a-Ortiz, J.P., Ireland, J., Zahniy, S., and Fleck, B.: 2017, {\it \aap} {\bf 606}, A10. \hyperlink{http://doi.org/10.1051/0004-6361/201730893 }{DOI}.

\bibitem[Odstrcil, Riley and Zhao(2004)]{Odstrcil-2004} Odstrcil, D., Riley, P., and Zhao, X.P.: 2004, {\it J.Geophys. Res. A} {\bf 109}, A02116. \hyperlink{http://doi.org/10.1016/j.jastp.2004.04.007}{DOI}. 

\bibitem[Pagano, Mackay and Yardley(2019a)]{Pagano-2019a}Pagano, P., Mackay, D.H., and Yardley, S.L.: 2019a, {\it The Astrophysical Journal} {\bf 883}, 112. \hyperlink{http://doi.org/10.3847/1538-4357/ab3e42}{DOI}.

\bibitem[Pagano, Mackay and Yardley(2019b)]{Pagano-2019b}Pagano, P., Mackay, D.H., and Yardley, S.L.: 2019b, {\it The Astrophysical Journal} {\bf 886}, 81. \hyperlink{http://doi.org/10.3847/1538-4357/ab4cf1}{DOI}.

\bibitem[Parker(1955)]{Parker-1955} Parker, E.N.: 1955, {\it \apj} {\bf 121}, 491. \hyperlink{http://doi.org/10.1086/146010}{DOI}.

\bibitem[Pesnell, Thompson and Chamberlin(2012)]{Pesnell-2012} Pesnell, W.D., Thompson, B.J., and Chamberlin, P.C.: 2012, {\it \solphys} {\bf 275}, 3.
\hyperlink{http://doi.org/10.1007/s11207-011-9841-3}{DOI}.

\bibitem[Pomoell, Lumme and Kilpua(2019)]{Pomoell-2019} Pomoell, J., Lumme, E., and Kilpua, E.: 2019, {\it Solar Physics} {\bf 294}, 41. \hyperlink{http://doi.org/10.1007/s11207-019-1430-x}{DOI}.


\bibitem[Rodkin \emph{et al.}(2017)]{Rodkin-2017} Rodkin, D., Goryaev, F., Pagano, P., Gibb, G., Slemzin, V., Shugay, Y., Veselovsky, I., and Mackay, D.H.: 2017, {\it Solar Phys.} {\bf 292}, 90. \hyperlink{http://doi.org/10.1007/s11207-017-1109-0}{DOI}.

\bibitem[Savcheva \emph{et al.}(2012)]{Savcheva-2012} Savcheva, A.S., Green, L.M., van Ballegooijen, A.A., and DeLuca, E.E.: 2012, {\it \apj.} {\bf 759}, 105. \hyperlink{http://doi.org/10.1007/s11207-011-9841-3}{DOI}.

\bibitem[Schou \emph{et al.}(2012)]{Schou-2012} Schou, J., Scherrer, P.H., Bush, R.I., Wachter, R., Couvidat, S., Rabello-Soares, M.C., Bogart, R.S., Hoeksema, J.T., Liu, Y., Duvall, T.L., Akin, D.J., Allard, B.A., Miles, J.W., Rairden, R., Shine R.A., Tarbell, T.D., Title, A.M., Wolfson,   C.J., Elmore D.F.,Norton A.A., and Tomczyk S.: 2012, {\it \solphys} {\bf 275}, 229.
\hyperlink{http://doi.org/10.1007/s11207-011-9842-2}{DOI}.

\bibitem[Schrijver \emph{et al.}(2006)]{Schrijver-2006} Schrijver, C.J., De Rosa, M.L., Metcalf, T.R., Liu, Y., McTiernan, J., R{\'e}gnier, S., Valori, G., Wheatland, M.S., and Wiegelmann, T.: 2006, {\it \solphys} {\bf 235}, 161.
\hyperlink{http://doi.org/10.1007/s11207-006-0068-7}{DOI}.

\bibitem[Spiegel and Zahn(1992)]{Spiegel-1992} Spiegel, E.A., and Zahn, J.-P.: 1992, {\it \aap} {\bf 265}, 106.

\bibitem[Strauss(1988)]{Strauss-1988} Strauss, H.R.: 1988, {\it \apj} {\bf 326}, 412.
\hyperlink{http://doi.org/10.1086/166104}{DOI}.

\bibitem[Su \emph{et al.}(2009)]{Su-2009} Su, Y., van Ballegooijen, A., Lites, B.W., Deluca, E.E., Golub, L., Grigis, P.C., Huang, G., and Ji, H.: 2009, {\it \apj} {\bf 691}, 105. \hyperlink{http://doi.org/10.1088/0004-637X/691/1/105}{DOI}.

\bibitem[SunPy Community \emph{et al.}(2015)]{SunPy-2015} SunPy Community, Mumford, S.J., Christe, S., P{\'e}rez-Su{\'a}rez, D., Ireland, J., Shih, A.Y., Inglis, A.R., Liedtke, S., Hewett, R.J., Mayer, F., Hughitt, K., Freij, N., Meszaros, T., and, ...: 2015, {\it Computational Science and Discovery} {\bf 8}, 14009. \hyperlink{http://doi.org/10.1088/1749-4699/8/1/014009}{DOI}.

\bibitem[T{\"o}r{\"o}k and Kliem(2005)]{Torok-2005} T{\"o}r{\"o}k, T., and Kliem, B.: 2005, {\it \apjl} {\bf 630}, L97. \hyperlink{http://doi.org/10.1086/462412}{DOI}.

\bibitem[Valori, Kliem and Keppens(2005)]{Valori-2005} Valori, G., Kliem, B., and Keppens, R.: 2005, {\it \aap} {\bf 433}, 335. \hyperlink{http://doi.org/10.1051/0004-6361:20042008}{DOI}.

\bibitem[van Ballegooijen(2004)]{vB-2004} van Ballegooijen, A.A.: 2004, {\it The Astrophysical Journal} {\bf 612}, 519. \hyperlink{http://doi.org/10.1086/422512}{DOI}.

\bibitem[van Ballegooijen and Mackay(2007)]{vB-2007} van Ballegooijen, A.A., and Mackay, D.H.: 2007, {\it \apj} {\bf 659}, 1713. \hyperlink{http://doi.org/10.1086/512849}{DOI}.

\bibitem[van Ballegooijen and Martens(1989)]{vB-1989} van Ballegooijen, A.A., and Martens, P.C.H.: 1989, {\it \apj} {\bf 343}, 971. \hyperlink{http://doi.org/10.1086/167766}{DOI}.

\bibitem[van Ballegooijen, Priest and Mackay(2000)]{vB-2000} van Ballegooijen, A.A., Priest, E.R., and Mackay, D.H.: 2000, {\it \apj} {\bf 539}, 983. \hyperlink{http://doi.org/10.1086/309265}{DOI}.

\bibitem[van Driel-Gesztelyi and Green(2015)]{vanDriel-2015} van Driel-Gesztelyi, L. and Green, L.M.: 2015, {\it Living Reviews in Solar Physics} {\bf 12}, 1. \hyperlink{http://doi.org/10.1007/lrsp-2015-1}{DOI}.

\bibitem[Wiegelmann and Sakurai(2012)]{Wiegelmann-2012} Wiegelmann, T., and Sakurai, T.: 2012, {\it Living Reviews in Solar Physics} {\bf 9}, 5. \hyperlink{http://doi.org/10.12942/lrsp-2012-5}{DOI}.

\bibitem[Yang, Sturrock and Antiochos(1986)]{Yang-1986} Yang, W.-H., Sturrock, P.A., and Antiochos, S.K.: 1986, {\it \apj} {\bf 309}, 383. \hyperlink{http://doi.org/10.1086/164610}{DOI}.


\bibitem[Yardley \emph{et al.}(2018a)]{Yardley-2018a} Yardley, S.L., Green, L.M., van Driel-Gesztelyi, L., Williams, D.R., and Mackay, D.H.: 2018a, {\it \apj} {\bf 866}, 8. \hyperlink{http://doi.org/10.3847/1538-4357/aade4a}{DOI}.

\bibitem[Yardley \emph{et al.}(2016)]{Yardley-2016}Yardley, S.L., Green, L.M., Williams, D.R., van Driel-Gesztelyi, L., Valori, G., and Dacie, S.: 2016, {\it The Astrophysical Journal} {\bf 827}, 151. \hyperlink{http://doi.org/10.3847/0004-637X/827/2/151}{DOI}.


\bibitem[Yardley, Mackay and Green(2018b)]{Yardley-2018b} Yardley, S.L., Mackay, D.H., and Green, L.M.: 2018b, {\it \apj} {\bf 852}, 82.
\hyperlink{http://doi.org/10.3847/1538-4357/aa9f20}{DOI}.


\bibitem[Yardley \emph{et al.}(2019)]{Yardley-2019}Yardley, S.L., Savcheva, A., Green, L.M., van Driel-Gesztelyi, L., Long, D., Williams, D.R., and Mackay, D.H.: 2019, {\it The Astrophysical Journal} {\bf 887}, 240.
\hyperlink{http://doi.org/10.3847/1538-4357/ab54d2}{DOI}.

\bibitem[Yeates, Mackay and van Ballegooijen(2007)]{Yeates-2007} Yeates, A.R., Mackay, D.H., and van Ballegooijen, A.A.: 2007, {\it \solphys} {\bf 245}, 87. \hyperlink{http://doi.org/10.1007/s11207-007-9013-7}{DOI}.


\bibitem[Zwaan(1985)]{Zwaan-1985} Zwaan, C.: 1985, {\it \solphys} {\bf 100}, 397. \hyperlink{http://doi.org/10.1007/BF00158438}{DOI}.





\end{thebibliography}
\end{document}